\documentclass[preprint,12pt,authoryear]{elsarticle}
\journal{Computer Physics Communications}
\usepackage{url}

\usepackage{pgfplots}
\usepackage{pgfplotstable}
\usepackage{tikz}
\usepackage{booktabs}
\usepackage{adjustbox}
\usetikzlibrary{arrows,shadows,backgrounds,calc,decorations.pathreplacing,decorations.pathmorphing}
\newcommand\QTfigure[6]   
{
	\draw
	($ (0 cm,0.5cm-.5*#2 cm) + (0 cm,-#5*.1 cm) +(#3cm,-#4cm)$) node[QT,minimum width=#2cm,draw]
	(Q#1) {$#1$}
	;
} 
\newcommand\QTCfigure[6]   
{
	\draw
	($ (1.7 cm,.5cm-.5*#2 cm) + (0 cm,-#5*.1 cm) +(#3cm,-#4cm)$) node[QTC,minimum width=#2cm,minimum height=5cm,draw]
	(Q#1) {$#1$}
	;
} 
\definecolor{webgreen}{rgb}{0,.5,0}
\definecolor{webblue}{rgb}{0,0,.8}
\definecolor{webred}{rgb}{0.8, 0, 0}   
\definecolor{webbrown}{rgb}{.6,0,0}
\definecolor{webyellow}{rgb}{0.98,0.92,0.73}
\definecolor{webgray}{rgb}{.753,.753,.753}
\newcommand{\gls}[1]{{\bfseries#1}}
\pgfplotscreateplotcyclelist{my color list}{%
	solid, color=webblue, every mark/.append style={solid, fill=webblue}, mark=*\\%
	densely dashdotted, color=webgreen, every mark/.append style={solid, fill=webred},mark=diamond*\\%
	densely dotted, color=webbrown, every mark/.append style={solid, fill=webgreen}, mark=triangle*\\%
	loosely dashed, color=webred, every mark/.append style={solid, fill=webbrown},mark=*\\%
	dotted, color=webblue, every mark/.append style={solid, fill=webyellow}, mark=square*\\%
	densely dotted, color=webgreen, every mark/.append style={solid, fill=gray}, mark=otimes*\\%
	dashed, color=webbrown, every mark/.append style={solid, fill=gray},mark=diamond*\\%
	densely dashed, every mark/.append style={solid, fill=gray},mark=square*\\%
	dashdotted, every mark/.append style={solid, fill=gray},mark=otimes*\\%
	dasdotdotted, every mark/.append style={solid},mark=star\\%
}

\begin{document}

\begin{frontmatter}


\title{How Amdahl's law restricts supercomputer applications\\and building ever bigger supercomputers}

\author[label1]{J\'anos V\'egh}
\ead{J.Vegh@uni-miskolc.hu}

\address[label1]{University of Miskolc, Hungary\\
       Department of Mechanical Engineering and Informatics\\
       3515 Miskolc-University Town, Hungary}

\begin{abstract}
This paper reinterprets Amdahl's law in terms of execution time and applies this simple model to supercomputing.
The systematic discussion results in a quantitative measure of computational efficiency of supercomputers and supercomputing applications, explains why
supercomputers have different efficiencies when using different benchmarks,
and why a new supercomputer intended to be the 1st on the TOP500 list
utilizes only 12~\% of its processors to achieve the 4th place only. 
Through separating non-parallelizable contribution to fractions
according to their origin,
Amdahl's law enables to derive a timeline for supercomputers
(quite similar to Moore's law) and 
describes why Amdahl's law limits the size of  supercomputers.
The paper validates that Amdahl's 50-years old model (with slight extension) correctly describes the performance limitations
of the present supercomputers.  
Using some simple and reasonable assumptions,  absolute performance bound of 
supercomputers is concluded, furthermore that serious enhancements are still necessary to achieve the exaFLOPS dream value.
\end{abstract}


\end{frontmatter}


\section{Introduction}

Supercomputers do have a quarter of century history for now, see~\cite{Top500:2016}.
The number of processors raised exponentially from the initial just-a-few processors, see~\cite{DongarraPerformance:1992}, to several millions, see~\cite{FuSunwaySystem2016}, and increased their computational performance (as well as electric power consumption) even more impressively. 
Supercomputers provide their  $R_{max}$ and $R_{peak}$
parameters when running a benchmark using all their processors, but
when needing not all available processors of a supercomputer and running programs other than benchmarks, they give not much hints for the effective computational performance for the case in question. 
Amdahl's law about  parallelly working systems introduces serious limitations on the joint performance,
so it is worth to scrutinize how Amdahl's law can assist in finding out execution time of a program as well as how it affected
the operational characteristics of supercomputers. 

As discussed by~\cite{ExponentialLawsComputing:2017}, "\textit{Moore's Law is one small component
	in an exponentially growing planetary
	computing ecosystem}". Using some simple assumptions,
they prove that from more closely Moore's law is described
by an S-curve rather than an infinite exponential.
Indeed, from the many components involved one can conclude different saturation points.
The fact that no more transistors can increase functionality of the present processors in a reasonable way,
caused a saturation about a decade ago (and started the age of multi-cores,
because density of transistors is not yet saturated).
Reaching the lithographic limits forced only introducing manufacturing tricks, but reaching the size of
atoms will surely cause a saturation,
so the industry is about to prepare for the post-Moore era, see~\cite{Post-Moore_2017}. 

Supercomputers with ever bigger computational power are planned.
As shown below, Amdahl's law represents another "\textit{small component in an exponentially growing planetary	computing ecosystem}".

\section{Amdahl's classic analyzis}\label{sec:amdahlclassic}

% % %  This is the stuff for the Amdahl's figure
\def\scalefact{0.8}
\tikzstyle{QT} = [top color=white, bottom color=blue!30,
minimum height={0.8},rotate=90,scale=\scalefact,xshift=-.5cm,yshift=-.5cm,
draw=blue!50!black!100, drop shadow] %, font=\small]
\tikzstyle{QTC} = [top color=green!80, bottom color=white,
minimum height={0.8},rotate=90,scale=\scalefact,xshift=-.5cm,yshift=-.8cm,
draw=blue!50!black!100, drop shadow] %, font=\small]
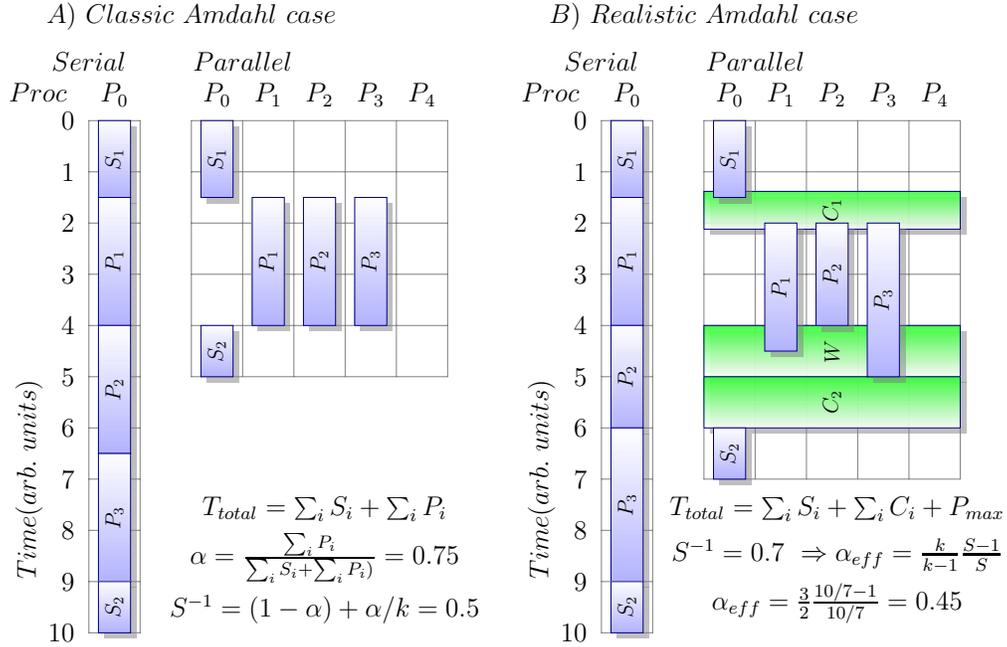
\begin{figure*} %[!b]
 \maxsizebox{\textwidth}{!}{
	\begin{tikzpicture}[scale=\scalefact,cap=round]
	
	% This is the initial figure, 
	\draw[style=help lines,step=1] (0,-10) grid (1,0);
	\node[right,above] at (2cm,1.6cm) {$A)\ Classic\ Amdahl\ case$};
	\node[right,above] at (0cm,.8cm) {$Serial$};
	\node[right,above] at (-1cm,.2cm) {$Proc$};
	\node[rotate=90,above=.5] at (-.7cm,-7cm) {$Time (arb.\ units)$};
	
	\foreach \x/\xtext in {0,...,0}
	\draw[xshift=\x cm+.5 cm,yshift=0.1cm]  node[above=.1]
	{$P_{\xtext}$};
	
	\foreach \y/\ytext in {0,...,10}
	\draw[yshift=-\y cm,xshift=-0.1cm]  node[left]
	{${\ytext}$};
	
	% These are the list of code fragments
	\QTfigure{S_1}{1.5}{0}{0}{0}{0}  % 
	\QTfigure{P_1}{2.5}{0}{0}{15}{0}  % 
	\QTfigure{P_2}{2.5}{0}{0}{40}{0}  % 
	\QTfigure{P_3}{2.5}{0}{0}{65}{0}  % 
	\QTfigure{S_2}{1}{0}{0}{90}{0}  % 
	
	%% This the parallelized version
	\draw[style=help lines,step=1] (2,-5) grid (7,0);
	\node[right,above] at (3cm,.8cm) {$Parallel$};
	\foreach \x/\xtext in {0,...,4}
	\draw[xshift=\x cm+2.5 cm,yshift=0.1cm]  node[above=.1]
	{$P_{\xtext}$};
	%
	%% These are the list of code fragments
	\QTfigure{S_1}{1.5}{2}{0}{0}{0}  % 
	\QTfigure{P_1}{2.5}{3}{0}{15}{0}  % 
	\QTfigure{P_2}{2.5}{4}{0}{15}{0}  % 
	\QTfigure{P_3}{2.5}{5}{0}{15}{0}  % 
	\QTfigure{S_2}{1}{2}{0}{40}{0}  % 
	
	\node[right,above] at (4.6cm,-8cm) {$T_{total}=\sum_i S_i + \sum_i P_i$};
	\node[right,above] at (4.6cm,-9.2cm) {$\alpha=\frac{\sum_iP_i} {\sum_i S_i + \sum_i P_i)} =0.75$};
	\node[right,above] at (4.6cm,-10cm) {$S^{-1}=(1-\alpha) +\alpha/k = 0.5$};
	
	% % This is the real case
	% This is the initial figure, 
	\draw[style=help lines,step=1] (10,-10) grid (11,0);
	\node[right,above] at (12cm,1.6cm) {$B)\ Realistic\ Amdahl\ case$};
	\node[right,above] at (10cm,.8cm) {$Serial$};
	\node[right,above] at (9cm,.2cm) {$Proc$};
	\node[rotate=90,above=.5] at (9.3cm,-7cm) {$Time (arb.\ units)$};
	
	\foreach \x/\xtext in {0,...,0}
	\draw[xshift=\x cm+10.5 cm,yshift=0.1cm]  node[above=.1]
	{$P_{\xtext}$};
	
	\foreach \y/\ytext in {0,...,10}
	\draw[yshift=-\y cm,xshift=-0.1cm+10cm]  node[left]
	{${\ytext}$};
	% These are the list of code fragments
	\QTfigure{S_1}{1.5}{10}{0}{0}{0}  % 
	\QTfigure{P_1}{2.5}{10}{0}{15}{0}  % 
	\QTfigure{P_2}{2.0}{10}{0}{40}{0}  % 
	\QTfigure{P_3}{3.0}{10}{0}{60}{0}  % 
	\QTfigure{S_2}{1}{10}{0}{90}{0}  % 
	
	%% This the parallelized version
	\draw[style=help lines,step=1] (12,-7) grid (17,0);
	\node[right,above] at (13cm,.8cm) {$Parallel$};
	\foreach \x/\xtext in {0,...,4}
	\draw[xshift=\x cm+12.5 cm,yshift=0.1cm]  node[above=.1]
	{$P_{\xtext}$};
	%
	%%% These are the list of code fragments
	\QTCfigure{C_1}{.5}{12}{0}{15}{0}  % 
	\QTCfigure{W}{1}{12}{0}{40}{0}  % 
	\QTCfigure{C_2}{1}{12}{0}{50}{0}  % 
	
	\QTfigure{S_1}{1.5}{12}{0}{0}{0}  % 
	\QTfigure{P_1}{2.5}{13}{0}{20}{0}  % 
	\QTfigure{P_2}{2.0}{14}{0}{20}{0}  % 
	\QTfigure{P_3}{3.0}{15}{0}{20}{0}  % 
	\QTfigure{S_2}{1}{12}{0}{60}{0}  % 

	\node[right,above] at (14.6cm,-8cm) {$T_{total}=\sum_i S_i + \sum_i C_i + P_{max}$};
	\node[right,above] at (14.6cm,-9cm) {$S^{-1}= 0.7\  \Rightarrow\alpha_{eff} = \frac{k}{k-1}\frac{S-1}{S} $};
	\node[right,above] at (14.6cm,-10cm) {$\alpha_{eff} =  \frac{3}{2}\frac{10/7 - 1}{10/7} = 0.45$};

	\end{tikzpicture}
}
	
	\caption{Illustrating Amdahl's law for idealistic and realistic cases.  For legend see text.}
	\label{fig:amdahl}
\end{figure*}

The most commonly known and cited limitation of the parallelization speedup (see~\cite{AmdahlSingleProcessor67}, the so called \emph{Amdahl's law}) 
is based on considering the fact that some parts ($P_i$) of a code can be parallelized,
some ($S_i$) must remain sequential.
It was early given by Amdahl, although the formula itself was constructed later by his successors.
Amdahl only wanted to draw the attention to the fact, that when putting together
several single processors, and using \gls{SPA}, the available speed gain due to using large-scale computing capabilities has a theoretical upper bound.
He also mentioned that data housekeeping (non-payload calculations)
causes some overhead, 
and  that \emph{the nature of that overhead appears to be sequential}.

The classic interpretation implies three\footnote{Another essential point which was missed by both~\cite{Karp:parallelperformance1990} and
	~\cite{ExponentialLawsComputing:2017},
	that \textit{the same computing model was used in all computers considered}.} 
essential restrictions,
but those restrictions are rarely mentioned (an exception:~\cite{Karp:parallelperformance1990}) in the textbooks on parallelization:
\begin{itemize}\setlength\itemsep{0em}
	\item the parallelized parts are of equal length in terms of execution time
	\item the housekeeping (controling the parallelization, passing parameters,
	waiting for termination, 
	exchanging messages, etc.) has no costs in terms of execution time
	\item the number of parallelizable chunks coincides with the number of available computing resources
\end{itemize}
\noindent Essentially,  this is why \emph{Amdahl's law represents a theoretical upper limit for parallelization gain}.
In Fig~\ref{fig:amdahl} the original process in the single-processor system comprises the sequential only parts $S_i$, and the parallelizable parts $P_i$. 
One can also see that the control components $C_i$ (not shown on the left side of the figure) are of the same nature as $S_i$,
the non-parallelizable components. This also means that even in the idealistic case
when $S_i$ are negligible, $C_i$ will create a bound for parallelization, \textit{independently of their origin}.

\subsection{Amdahl's case under realistic conditions}
\label{sec:AmdahlRealistic}

The realistic case (shown in the right side of Fig~\ref{fig:amdahl}) is, however,  that
the parallelized parts are \emph{not} of equal length (even if they contain exactly the same 
instructions), the hardware operation in modern processors may execute them in 
considerably different times; for examples see \cite{Vegh:2014:ICSOFTsemaphore} and \cite{HennessyArchitecture2007}, and references cited therein;
the operation of hardware accelerators inside a core; the network operation between processors; etc.
One can also see  that the time required to control parallelization is not negligible and varying.
This represents another performance bound.

The figure also calls the attention to that the static correspondence between program chunks and
processing units can be very inefficient: all assigned processing units must wait the delayed unit
and also some capacity is lost if the number of computing resources exceeds the number of parallelized chunks.

It is much worse, if the number of the processing units is smaller than that of parallelized threads:
the processor must organize severals "rounds" for remaining  threads,
with all disadvantages of the duty of synchronization, see~\cite{YavitsMulticoreAmdahl2014,SynchronizationEverything2013}.
Also, the longer code chunks $P_i$ are, the higher is the chance to waste computing
capacity of the processing units which already finished their task. 
Note that here the programmer (the person or the compiler) has to organize the job, in the spirit of
\textit{Single Processor Approach} (SPA): at some point the single processor splits the 
execution, transmits necessary parameters to some other processing units, starts
their processing, then waits termination of the started processings.
Real-life programs show sequential-parallel behavior, with variable degree
of parallelization, see~\cite{YavitsMulticoreAmdahl2014} and even apparently 
massively parallel algorithms change their behavior during processing, see~\cite{Pingali:2011:TaoOfParallelism}.
\index{parallelism!real life}
All these make Amdahl's original mode non-applicable.

\subsection{Factors affecting parallelism}

Usually,  Amdahl's law is expressed with the formula 

\begin{equation}
S^{-1}=(1-\alpha) +\alpha/k \label{eq:AmdahlBase}
\end{equation}

\noindent where $k$ is the number of parallelized code fragments, 
$\alpha$ is the ratio of the parallelizable part  to total,
$S$ is the measurable speedup. 
The assumption can be visualized that (assuming many processors)
in $\alpha$ fraction of the running time processors are processing data,
in (1-$\alpha$) fraction they are waiting (all but one). That is $\alpha$ describes
how much, in average, the processors are utilized.
Having those data, the resulting speedup can be estimated. 

A general misconception (introduced by followers of Amdahl) is to assume that Amdahl's law
is valid for software only and that $\alpha$ contains
something like ratio of numbers of the corresponding instructions.
Actually,\textit{ Amdahl's law is valid for any partly parallelizable activity (including computer unrelated ones) and $\alpha$  is given as the ratio of the time spent with parallelizable activity 
	to total time}.

For a system under test, where  $\alpha$ is not \textit{a priory} known,
one can derive from the measurable speedup  $S$ 
an \emph{effective parallelization} factor as

\begin{equation}
\alpha_{eff} = \frac{k}{k-1}\frac{S-1}{S} \label{equ:alphaeff}
\end{equation}
\index{parallelism!effective}
\index{effective parallelism}

\noindent Obviously, this is not more than $\alpha$ expressed in terms of $S$ and $k$ from Equ.~(\ref{eq:AmdahlBase}).
So, for the classical case, $\alpha = \alpha_{eff}$; which simply means that
in \emph{ideal} case the actually measurable effective parallelization 
achieves the theoretically possible one.
In other words, $\alpha$ describes a system the \emph{architecture} of which is completely known,
while $\alpha_{eff}$ characterizes a system the \emph{performance} of which is known from experiments.
Again in other words,  $\alpha$ is the \emph{theoretical upper limit}, which can hardly be achieved,
while $\alpha_{eff}$ is the \emph{experimental actual value}, which describes the complex architecture and the actual conditions. 
%It is worth to note, that $\alpha_{eff}$ is an 
%absolute measure of utilizing the available parallel processing capacity, see section \ref{sec:practical}.

$\alpha_{eff}$ can then be used to refer back to Amdahl's classical
assumption even in the realistic case when parallelized chunks 
have different lengths and the overhead to organize parallelization is not negligible.
Speedup $S$  can be measured and $\alpha_{eff} $ can be utilized
to characterize the measurement setup and conditions, 
how much from the theoretically possible maximum parallelization is realized.

Note that in the case of real tasks a kind of Sequential/Parallel Execution Model,  see~\cite{YavitsMulticoreAmdahl2014}, shall be applied, which cannot use 
the simple picture reflected by $\alpha$, but $\alpha_{eff}$ gives a good merit
of degree of parallelization for duration of execution of the process on the given hardware configuration,
and can be compared to results of technology-dependent parametrized formulas.
Numerically ($1-\alpha_{eff}$) equals with the $f$ value, established theoretically by~\cite{Karp:parallelperformance1990}.
\index{parallelism!supercomputer}
In the case of supercomputer HW, it can measure what
fraction of running time spend processors with non-payload activity.

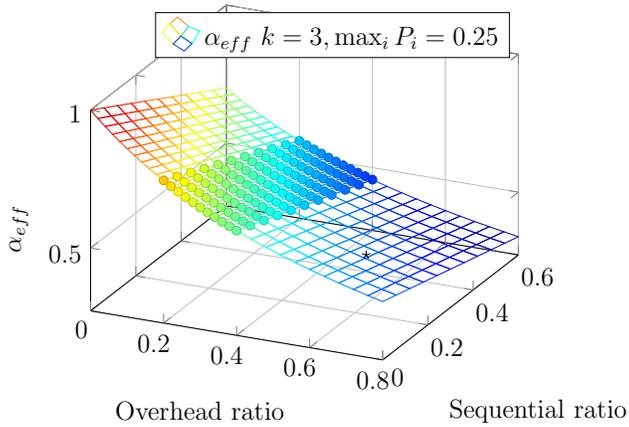
\begin{figure}
	%	\maxsizebox{\columnwidth}{!}
	%	{
	\begin{tikzpicture}  [scale=0.83]%[scale=.83]
	\begin{axis}[
	ylabel=Sequential ratio ,
	xlabel=Overhead ratio, 
	zlabel=$\alpha_{eff}$,
	zmax=1,
	grid=both,
	colormap/jet,
	%colormap/cool
	]
	\addplot3[
	mesh, 
	samples=11,
	domain=0:0.2,y domain=0:0.6,
	]
	{ 
		(3./(3-1))* % % k is the number of parallelized threads
		( 1.- ((x+.25*(1.+y))/(x+3.*.25)))
	};
	\addplot3[
	mesh, scatter,
	samples=11,
	domain=0.2:0.4,y domain=0:0.6,
	]
	{ 
		(3./(3-1))* % % k is the number of parallelized threads
		( 1.- ((x+.25*(1.+y))/(x+3.*.25)))
	};
	\addplot3[
	mesh, 
	%	surf,faceted color=blue,
	samples=11,
	domain=0.4:0.8,y domain=0:0.6,
	]
	{ 
		(3./(3-1))* % % k is the number of parallelized threads
		( 1.- ((x+0.25*(1.+y))/(x+3.*0.25)))
	};
	\addlegendentry{$\alpha_{eff}\ k=3, \max_i P_i=0.25$}
	
	\addplot3+[only marks] coordinates {
		(0.6,0.25, 0.45) };
	\end{axis}
	\end{tikzpicture}
	%	}
	\caption{Behavior of the effective parallelization $\alpha_{eff}$ in function of overhead ratio (compared to parallelizable payload execution length) and ratio of sequential part (compared to total sequential execution time).}
	\label{EffectiveFactor}
\end{figure}

\subsection{Applying $\alpha_{eff}$ to Amdahl's model}
\label{sec:decreasing control time}
With our notations, in the classical Amdahl case on the left side in Fig.~\ref{fig:amdahl}

\begin{equation}
S= \frac{\sum_i S_i  + \sum_i P_i}{\sum_i S_i + \max_i P_i} = 2
\end{equation}

\noindent and 

\begin{equation}
\alpha = \alpha_{eff}= \frac{\sum_i Pi}{\sum_i S_i+\sum_i Pi} = 3/4
\end{equation}

Now we can compare effective parallelizations in the two cases shown in Fig.~\ref{fig:amdahl}.
%In the classical case $S=2$ and so $\alpha_{eff}=0.75$. 
In the realistical case $S =10/7$, which results in 

\begin{equation}
\alpha_{eff}= \frac{3}{2}\frac{10/7 - 1}{10/7} = 0.45
\end{equation}

\noindent The overhead and the different duration of the parallelized parts
reduced the effective parallelization drastically compared to the theoretically achievable value. 
Fig~\ref{EffectiveFactor} gives a feeling on the effect of computer system behaviour
on effective parallelization.  The middle region (marked by balls) is mentioned by Amdahl as typical range of overhead. The asterisk  in the figure shows the "working point"
corresponding to the values used in Fig~\ref{fig:amdahl}. 

One can see that effective parallelization drops quickly with both increasing 
overhead and sequential parts of the program. This fact draws the attention to the idea that
\emph{through 
	decreasing either the control time or the sequential-only fraction of the code (or both),
	and utilizing
	the otherwise wasted processing capacity, a serious gain in the effective parallelization
	can be reached}.

\section{Efficiency of parallelization}
%\section{Consequences of Amdahl's model}
The distinguished constituent in Amdahl's classic analysis is parallelizable payload fraction $\alpha$,
all the rest (including wait time, communication, system contribution and any other non-payload activity) goes into the "sequential-only" fraction  according to this extremely simple model.

When using several processors, one of them makes the sequential-only calculation, the others are waiting
(use the same amount of time). So, when calculating the speedup, one calculates

\begin{equation}
S=\frac{(1-\alpha)+\alpha}{(1-\alpha)+\alpha/k} =\frac{k}{k(1-\alpha)+\alpha}
\end{equation}
hence the  \textit{efficiency}\footnote{This quantity is almost exclusively used to describe computing performance of multi-processor systems.
	In the case of supercomputers, $\frac{R_{max}}{R_{peak}}$ is provided, which is identical with $E$} (how speedup scales with the number of processors)
\begin{equation}
E = \frac{S}{k}=\frac{1}{k(1-\alpha)+\alpha}\label{eq:soverk}
\end{equation}

\index{efficiency of parallelization}
\index{speedup due to parallelization}

\begin{figure*}
\maxsizebox{\textwidth}{!}{
		\begin{tabular}{rr}
		\maxsizebox{\columnwidth}{!}
		{
			\begin{tikzpicture}
			\begin{axis}[%
			legend style={
				cells={anchor=west},
				legend pos={south east},
			},
			cycle list name={my color list},
			xmin=0, xmax=51,% x scale
			ymode=log,
			ymin=1e-5, ymax=5e-3, % y scale
			xlabel=Rank of supercomputer in 2000,
			ylabel=$(1-\alpha_{eff})$ ,
			scatter/classes={%
				MPP={ mark=diamond*,  draw=webgreen}, Cluster={ mark=triangle*, draw=webblue}, 
				%			MPPR={ mark=square*,  draw=webgreen}, ClusterR={ mark=triangle*, draw=webblue},
				SCMCS={mark=triangle*,draw=webred} }
			]
			\addplot[scatter,only marks, %
			scatter src=explicit symbolic]%
			table[meta=label,    /pgf/number format/read comma as period
			] {
				x y label
				1	3,614E-05	MPP
				2	1,375E-04	MPP
				3	1,482E-04	MPP
				4	3,103E-04	MPP
				5	2,690E-03	MPP
				6	3,117E-03	MPP
				7	4,247E-04	MPP
				8	4,247E-04	MPP
				9	1,362E-03	MPP
				10	3,493E-04	MPP
				11	6,552E-04	MPP
				12	2,355E-04	MPP
				13	4,112E-04	MPP
				14	5,575E-04	MPP
				15	5,575E-04	MPP
				16	5,811E-04	MPP
				17	4,201E-03	MPP
				18	4,894E-04	MPP
				19	7,143E-04	MPP
				20	7,102E-04	MPP
				21	9,167E-04	MPP
				22	9,167E-04	MPP
				25	1,685E-03	MPP
				26	7,362E-04	MPP
				27	6,746E-04	MPP
				28	2,227E-03	MPP
				29	6,005E-04	MPP
				30	8,343E-04	MPP
				31	8,343E-04	MPP
				32	8,343E-04	MPP
				33	8,343E-04	MPP
				34	5,828E-04	MPP
				35	3,267E-04	MPP
				36	4,592E-04	MPP
				37	9,823E-04	MPP
				38	1,551E-03	MPP
				39	7,889E-04	MPP
				40	7,889E-04	MPP
				41	1,120E-03	MPP
				42	1,136E-03	MPP
				43	1,500E-03	MPP
				44	1,282E-03	MPP
				45	9,241E-04	MPP
				46	1,352E-03	MPP
				47	7,905E-04	MPP
				49	1,369E-03	MPP
			};
			
			\addplot[scatter,only marks, mark=triangle*,
			scatter src=explicit symbolic]%
			table[meta=label, /pgf/number format/read comma as period] {
				x y label
				23	6,749E-04	Cluster
				24	6,749E-04	Cluster
				50	1,063E-03	Cluster
			};
			
			\addplot+[ mark=diamond,  draw=webgreen] table[y={create col/linear regression={y=Y}},% mark=rectangle*,
			meta=label,    /pgf/number format/read comma as period
			] {
				x Y label
				1	3,614E-05	MPPR
				2	1,375E-04	MPPR
				3	1,482E-04	MPP
				4	3,103E-04	MPP
				5	2,690E-03	MPP
				6	3,117E-03	MPP
				7	4,247E-04	MPP
				8	4,247E-04	MPP
				9	1,362E-03	MPP
				10	3,493E-04	MPP
				11	6,552E-04	MPP
				12	2,355E-04	MPP
				13	4,112E-04	MPP
				14	5,575E-04	MPP
				15	5,575E-04	MPP
				16	5,811E-04	MPP
				17	4,201E-03	MPP
				18	4,894E-04	MPP
				19	7,143E-04	MPP
				20	7,102E-04	MPP
				21	9,167E-04	MPP
				22	9,167E-04	MPP
				25	1,685E-03	MPP
				26	7,362E-04	MPP
				27	6,746E-04	MPP
				28	2,227E-03	MPP
				29	6,005E-04	MPP
				30	8,343E-04	MPP
				31	8,343E-04	MPP
				32	8,343E-04	MPP
				33	8,343E-04	MPP
				34	5,828E-04	MPP
				35	3,267E-04	MPP
				36	4,592E-04	MPP
				37	9,823E-04	MPP
				38	1,551E-03	MPP
				39	7,889E-04	MPP
				40	7,889E-04	MPP
				41	1,120E-03	MPP
				42	1,136E-03	MPP
				43	1,500E-03	MPP
				44	1,282E-03	MPP
				45	9,241E-04	MPP
				46	1,352E-03	MPP
				47	7,905E-04	MPP
				49	1,369E-03	MPPR
			};
			
			\addplot+[ mark=triangle, draw=webblue] table[y={create col/linear regression={y=Y}}, mark=diamond*,
			meta=label,    /pgf/number format/read comma as period
			] {
				x Y label
				23	6,749E-04	Cluster
				24	6,749E-04	Cluster
				50	1,063E-03	Cluster
			};
			\addlegendentry{MPP in 2000}
			\addlegendentry{Cluster in 2000}
			\addlegendentry{Regression of MPP in 2000}
			\addlegendentry{Regression of cluster in 2000}
			\end{axis}
			\end{tikzpicture}
		}
		&
		\maxsizebox{\columnwidth}{!}
		{
			\begin{tikzpicture}
			\begin{axis}[%
			legend style={
				cells={anchor=west},
				legend pos={south east},
			},
			cycle list name={my color list},
			xmin=0, xmax=51,% x scale
			ymode=log,
			ymin=5e-8, ymax=2e-5, % y scale
			xlabel=Rank of supercomputer in 2016,
			ylabel=$(1-\alpha_{eff})$,
			scatter/classes={%
				MPP={ mark=diamond*,  draw=webgreen}, Cluster={ mark=triangle*, draw=webblue}, 
				MPPR={ mark=square*,  draw=webgreen}, ClusterR={ mark=triangle*, draw=webblue},
				SCMCS={mark=triangle*,draw=webred} }
			]
			\addplot[scatter,only marks,%
			scatter src=explicit symbolic]%
			table[meta=label] {
				x y label
				3	9.656E-07	MPPR
				4	1.096E-07	MPP
				6	2.191E-07	MPP
				7	1.221E-06	MPP
				8	2.087E-06	MPP
				9	1.689E-06	MPP
				10	1.560E-06	MPP
				13	3.756E-07	MPP
				14	4.383E-07	MPP
				16	2.250E-06	MPP
				17	6.107E-07	MPP
				18	6.107E-07	MPP
				24	2.446E-06	MPP
				29	8.635E-07	MPP
				30	8.635E-07	MPP
				32	4.464E-06	MPP
				35	4.815E-06	MPP
				36	2.997E-06	MPP
				37	2.997E-06	MPP
				45	1.052E-06	MPP
				49	4.131E-06	MPP
				50	4.682E-06	MPP
			};
			
			\addplot[scatter,only marks,%
			scatter src=explicit symbolic]%
			table[meta=label] {
				x y label
				2	1.991E-07	ClusterR
				5	1.040E-07	Cluster
				11	1.225E-06	Cluster
				12	1.402E-06	Cluster
				15	1.163E-06	Cluster
				19	9.811E-06	Cluster
				20	9.811E-06	Cluster
				21	3.036E-06	Cluster
				22	6.173E-06	Cluster
				23	9.318E-07	Cluster
				25	5.204E-06	Cluster
				26	1.246E-06	Cluster
				27	6.743E-07	Cluster
				28	3.160E-06	Cluster
				31	1.365E-05	Cluster
				33	2.677E-06	Cluster
				34	1.718E-06	Cluster
				38	1.243E-06	Cluster
				39	4.628E-06	Cluster
				40	2.347E-06	Cluster
				41	9.587E-06	Cluster
				42	2.772E-06	Cluster
				43	4.132E-06	Cluster
				44	5.438E-06	Cluster
				46	2.976E-06	Cluster
				47	6.123E-06	Cluster
				48	7.291E-06	Cluster
			};
			
			\addplot+[ mark=diamond,  draw=webgreen] table[y={create col/linear regression={y=Y}}, %mark=rectangle*,
			meta=label,    /pgf/number format/read comma as period
			]
			{
				x Y label
				3	9.656E-07	MPP
				4	1.096E-07	MPP
				6	2.191E-07	MPP
				7	1.221E-06	MPP
				8	2.087E-06	MPP
				9	1.689E-06	MPP
				10	1.560E-06	MPP
				13	3.756E-07	MPP
				14	4.383E-07	MPP
				16	2.250E-06	MPP
				17	6.107E-07	MPP
				18	6.107E-07	MPP
				24	2.446E-06	MPP
				29	8.635E-07	MPP
				30	8.635E-07	MPP
				32	4.464E-06	MPP
				35	4.815E-06	MPP
				36	2.997E-06	MPP
				37	2.997E-06	MPP
				45	1.052E-06	MPP
				49	4.131E-06	MPP
				50	4.682E-06	MPP
			};
			
			\addplot+[ mark=triangle, draw=webblue] table[y={create col/linear regression={y=Y}}, %mark=o*,
			meta=label,    /pgf/number format/read comma as period
			]
			{
				x Y label
				2	1.991E-07	Cluster
				5	1.040E-07	Cluster
				11	1.225E-06	Cluster
				12	1.402E-06	Cluster
				15	1.163E-06	Cluster
				19	9.811E-06	Cluster
				20	9.811E-06	Cluster
				21	3.036E-06	Cluster
				22	6.173E-06	Cluster
				23	9.318E-07	Cluster
				25	5.204E-06	Cluster
				26	1.246E-06	Cluster
				27	6.743E-07	Cluster
				28	3.160E-06	Cluster
				31	1.365E-05	Cluster
				33	2.677E-06	Cluster
				34	1.718E-06	Cluster
				38	1.243E-06	Cluster
				39	4.628E-06	Cluster
				40	2.347E-06	Cluster
				41	9.587E-06	Cluster
				42	2.772E-06	Cluster
				43	4.132E-06	Cluster
				44	5.438E-06	Cluster
				46	2.976E-06	Cluster
				47	6.123E-06	Cluster
				48	7.291E-06	Cluster
			};
			\addlegendentry{MPP in 2016}
			\addlegendentry{Cluster in 2016}
			\addlegendentry{Regression of MPP in 2016}
			\addlegendentry{Regression of cluster in 2016}
			\end{axis}
			\end{tikzpicture}
		}	
		\\
	\end{tabular}
}
	\caption{Dependence of $(1-\alpha_{eff})$ on architectural solution of supercomputer in 2000 and 2016.  Data derived using the $HPL$ benchmark}
	\label{fig:SCarchitecture}
\end{figure*}
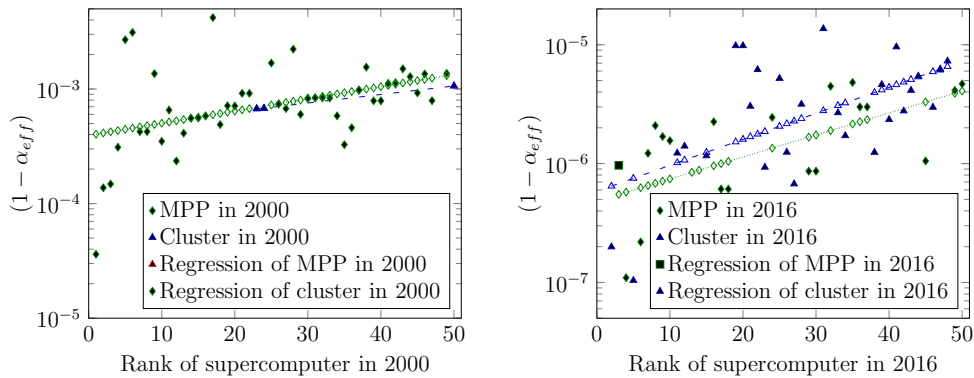

\begin{table}[ht]
	{
		\small\sf\centering
		\caption{Performance of Various Computers Using Standard Linear Equations ($HPL$) Software at the beginning of the supercomputer age %\protect{\cite{DongarraPerformance:1992}}
		}
		\label{tab:Dongarra1992}
		\begin{tabular} 
			{lrrr} %
			\toprule
			Computer Model    & N proc & Efficiency  & $1-\alpha_{eff}$  \\
			\midrule
			Cray Y-MP C90 &	16 &	0.69 &	2.995E-02 \\
			NEC SX-3      &  	2  &	0.91 &	9.890E-02 \\
			Cray Y-MP/8   &	8  &	0.87 &	2.135E-02 \\
			Fujitsu AP 1000 &	512 &	0.29 &	4.791E-03 \\
			IBM 3090/600S VF &	6	& 0.94	 & 1.277E-02 \\
			Intel Delta   &	512	& 0.03	 & 6.327E-02 \\
			Alliant FX/2800-200 &	14 &	0.79 &	2.045E-02 \\
			NCUBE/2       &	1024 &	0.12 &	7.168E-03 \\
			Convex C3240	 & 4	& 0.95 &	1.754E-02 \\
			Parsytec FT-400	 &	400	& 0.55 &	2.051E-03 \\
			\bottomrule
		\end{tabular}
	}
\end{table}

The importance of producing more efficient parallelization 
manifests  from the very beginnings, see Table~\ref{tab:Dongarra1992}. It could be observed that 
supercomputers having an order of magnitude higher number of
processors should have an order of magnitude lower value of $(1-\alpha_{eff})$  to produce reasonable efficiency.
Ironically enough, the higher number of cores is accompanied with lower efficiency, but at the same time, with better  ($1-\alpha_{eff}$) value.
It looks like they did seriously consider Amdahl's  very reasonable conclusion:"
\emph{the effort expended
	on achieving high parallel processing rates is wasted unless it is accompanied by achievements in
	sequential processing rates of very nearly the same magnitude}"~\cite{AmdahlSingleProcessor67}.
The one (Intel Delta) which did not enhance architecture,
had tragically low efficiency.

%\MEtikzfigure[wide]{
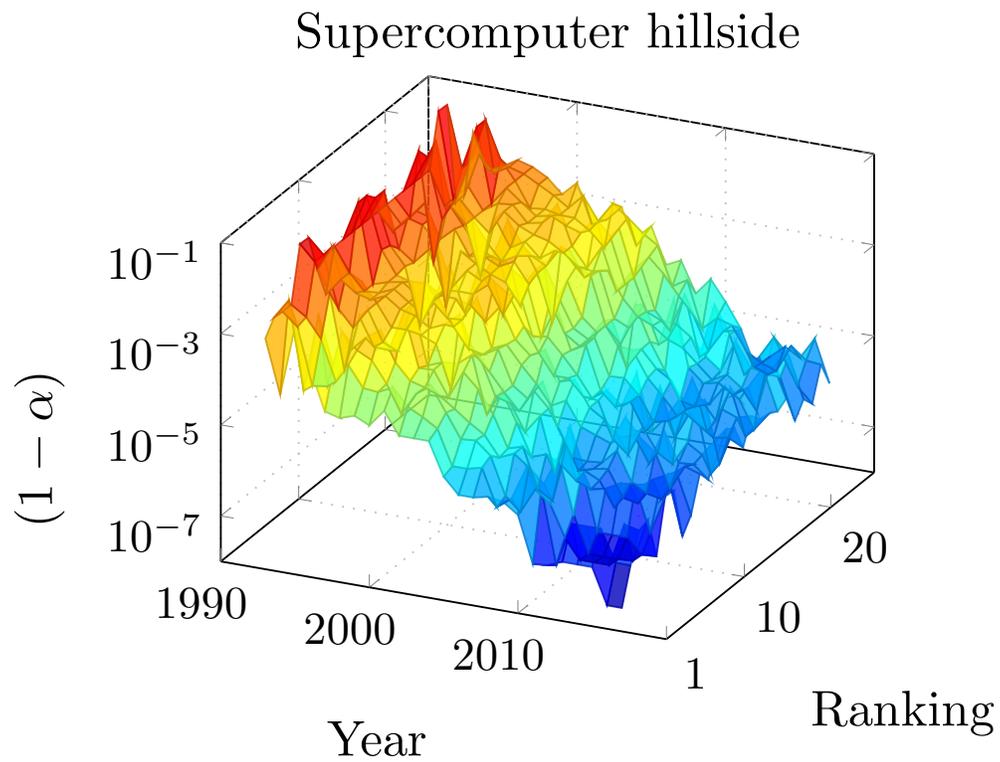
\begin{figure*}
	\maxsizebox{\textwidth}{!}
	{
	\begin{tikzpicture}[scale=2.4]
	\pgfplotstableset{%
		col sep=semicolon,
		z index=0,
		y index=1,
		x index=2,
		header=false
	}%
	\pgfplotsset{
		%   width=15cm,compat=1.8,
		every axis/.append style={
			scale only axis,
			width=\textwidth,
			height=\textwidth,
			xtick={1990,2000,2010},
			ytick={1,10,20},
			ztick={1e-7,1e-5,1e-3,1e-1}
		},
		/tikz/every picture/.append style={
			trim axis left,
			trim axis right,
		}
	}
	\begin{axis}
	[ %nodes near coords={(\coordindex)},
	domain=1980:2020,
	domain y=1:40,
	footnotesize,
	title={Supercomputer hillside},
	/pgf/number format/.cd,
	use comma,
	1000 sep={},
	xlabel=Year,
	ylabel=Ranking,
	zlabel=$(1-\alpha)$,
	xmin=1990, xmax=2020,% x scale
	ymin=1, ymax=25, % y scale
	zmin=1e-8, zmax=1e-1, % z scale
	xlabel=Year,
	zmode = log,
	mesh/cols=25,
	grid=major,
	grid style={dotted},
	colormap/jet,
	zmode=log,
	]
	
	\addplot3[%
	surf,%shader=interp,
	opacity=0.8
	] table {SupercomputerHillsideNew.csv};
	
	\end{axis}
	\end{tikzpicture}
}
	\caption{The Top500 supercomputer parallelization efficiency.
		The ($1-\alpha_{eff}$) parameter for the past 25 years and the  (by $R_{max}$) first 25 computers. Data derived using the \gls{HPL} benchmark.}
	\label{SupercomputerHillside}
\end{figure*}

\section{Applying $\alpha_{eff}$ to supercomputers}

Comparing supercomputers having different architectures,
processors, manufacturers, number of processors, technological age, etc. is not easy at all.
In order to make different \textit{architectures} comparable, with as low distortion as possible, 
the same benchmark program \gls{HPL} is used to qualify different supercomputers, since the beginning of supercomputer era.
The distortion caused by the measurement device (the benchmark program) cannot be eliminated,
but using the same benchmark causes the same distortion,
so at least makes makes performance of different supercomputers comparable.

It should be noticed (see Equ.~(\ref{eq:soverk})) that $\frac{1}{E}$ is a linear function of the number of processors, and its slope equals to $(1-\alpha_\Delta)$.
This can be used to quantitize how changes affecting any of the 
contributions to the non-parallelizable part influence $\alpha_{eff}$.
The numerical value calculated in this way is quite near  to the value  calculated using Equ.~(\ref{equ:alphaeff}),
and so it is not displayed in the figures. 

\subsection{Architecture}

$\alpha_{eff}$ can also be utilized during supercomputer building, measuring execution times of the benchmark program using two different number of processors. 
One can estimate value of $\alpha_\Delta$ even for the intermediate regions,
i.e. without knowing the execution time on a single processor
(from technical reasons, it is the usual case for supercomputers).
From the value of  $\alpha_\Delta$ the efficiency 
of the planned supercomputer can be calculated\footnote{This of course assumes that $\alpha_{eff}$ is independent of the number of processors, which seems to be valid at this level of approximation.}, and so
from a handful of processors one can find out if the 
supercomputer under construction will beat the No. 1 in~\cite{Top500:2016}.
It would be reasonable to consider the experience: 
"\emph{Virtually every practicing computer architect knows Amdahl's Law. Despite this,
	we almost all occasionally expend tremendous effort optimizing some feature
	before we measure its usage. Only when the overall speedup is disappointing do
	we recall that we should have measured first before we spent so much effort
	enhancing it!"}~\cite{HennessyArchitecture2007}

\subsection{Supercomputer timeline}
%\subsection{Characterizing supercomputers}

%\MEtikzfigure[wide]{
\begin{figure*}
	\begin{tikzpicture}[scale=.95]
	\begin{axis}
	[
	title={Supercomputers, Top 500 1st-3rd},
	width=\textwidth,
	cycle list name={my color list},
	legend style={
		cells={anchor=east},
		legend pos={north east},
	},
	xmin=1993, xmax=2017,% x scale
	ymin=1e-8, ymax=1e-2, % y scale
	xlabel=Year,
	/pgf/number format/1000 sep={},
	ylabel=$(1-\alpha)$,
	ymode=log,
	log basis x=2,
	]
	\addplot table [x=a, y=b, col sep=comma] {Top500-0.csv};
	\addlegendentry{$1st $}
	\addplot table [x=a, y=c, col sep=comma] {Top500-0.csv};
	\addlegendentry{$2nd $}
	%   \addplot table[ x=a, %y=e,
	%    y={create col/linear regression={x=a,y=c}}, 
	%     col sep=comma] % compute a 
	%    {dat/Top500-0.csv};		
	%		\addlegendentry{Regression of 2nd by $\ R_{Max}$}
	\addplot table [x=a, y=d, col sep=comma] {Top500-0.csv};
	\addlegendentry{$3rd $}
	\addplot table [x=a, y=e, col sep=comma] {Top500-0.csv};
	\addlegendentry{$Best\ \alpha$}
	%		\addplot[thick, color=webbrown] plot coordinates {
	%			(1993, 1.5e-3)  
	%			(2017,.7e-7) 
	%		};
	%		\addlegendentry{Trend of $(1-\alpha)$}
	
	\addplot table[ x=a, %y=e,
	y={create col/linear regression={x=a,y=b}}, 
	col sep=comma] % compute a 
	{Top500-0.csv};		
	\addlegendentry{Regression of best $\ R_{Max}$}
	\addplot table[ x=a, %y=e,
	y={create col/linear regression={x=a,y=e}}, 
	col sep=comma] % compute a 
	{Top500-0.csv};		
	\addlegendentry{Regression of best $\ \alpha$}
	\addplot[only marks, color=red, mark=star,  mark size=3, very thick] plot coordinates {
		(2016,33e-9) 
	};
	\addlegendentry{TaihuLight}
	\end{axis}
	\end{tikzpicture}
	\caption{The trend of development of ($1-\alpha_{eff}$) in the past 25 years, based on the 
		first three (by $R_{max}$) and the first (by ($1-\alpha_{eff}$)) supercomputers in the year in question. Data derived using the \gls{HPL} benchmark.}
	\label{SupercomputerTimeline}
\end{figure*}
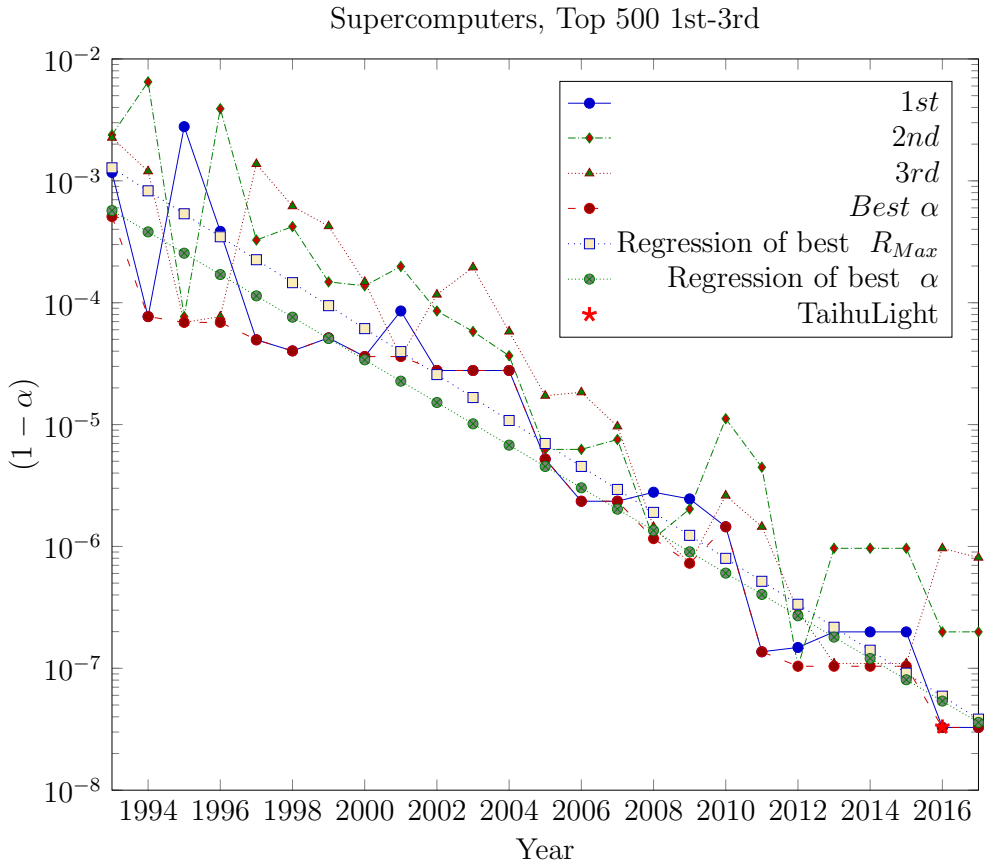

In supercomputers, the "(apparently) sequential part" is technically of different origin -- and as will be shown below, 
can be separated to contributions of different nature --, but has the same effect on ($1-\alpha_{eff}$) in the frame of the classic model.
To scrutinize a possible dependency of ($1-\alpha_{eff}$) on time, 
data  covering 25 years of "supercomputer age", see~\cite{Top500:2016}, 
has been analyzed. The ratio of data  $R_{max}$ and $R_{peak}$ provided $E$,  and using Equ.~(\ref{eq:soverk}), 
($1-\alpha_{eff}$) has been calculated in function of time and ranking, see Fig.~\ref{SupercomputerHillside}. It looks like ($1-\alpha_{eff}$)
changes in an exponential-like way, both with time and with ranking in a given year.

In Fig.~\ref{SupercomputerTimeline}
$(1-\alpha_{eff})$ values
for the top 3 supercomputers (ranked by $R_{max}$) are displayed  in function of time.
The figure also contains diagram of the best (ranked by $(1-\alpha_{eff})$) computer in the year,
which confirms that high computing performance strongly correlates with 
efficiency of parallelization.
Both methods lead to the same trend: the two regression lines shown in the figure are calculated for the \#1 by $R_{Max}$ and by $(1-\alpha)$, respectively.
It looks like this development path
(independently of technology, manufacturer, number and type of processors, etc.) shows a semi-logarithmic behavior over a quarter century.
It is able to forecast expected behavior of performance in the coming years,  in the same way as Moore observation does  for single processors.

One might think it is another appearance of Moore's law.
However, consider that when calculating $\frac{R_{Max}}{R_{Peak}}$,  single-processor 
performance (clock frequency and component density, etc.)  consequences of the Moore-observation are removed. What remains: how perfectly  single-processors can work together.

For the first look it might be surprising to look for any dependency  of $(1-\alpha_{eff})$ on 
year of construction.
Consider, however, that \textit{it represents the non-parallellizable fraction of execution time}, and the need for higher performance requires to increase the number of processors comprised.
The key is Eq.~(\ref{eq:soverk}):
$R_{Peak}$  increases linearly with $k$, and efficiency decreases according to $\frac{1}{k(1-\alpha_{eff})}$.
Supercomputers need a large number of processors, so the only way
to put them together in an efficient and economic way, is to decrease  $(1-\alpha_{eff})$.
Moore's law assures that in the same year a very similar \gls{SPA} technology is used
by all manufacturers, so that computers from different manufacturers can be compared.
Amdahl's law assures, that the only way to build  many-processor
system with higher $R_{Max}$ from the same single-processor components,
is to reduce  $(1-\alpha_{eff})$.

The architectural solution does not play a very important role here. As Fig.~\ref{fig:SCarchitecture} depicts,
both major architectural technologies result  $(1-\alpha_{eff})$ in the same order of magnitude,
and $(1-\alpha_{eff})$ raises with ranking of the computer, so some other factors may decide on  ranking.

\begin{figure*}
	\maxsizebox{\textwidth}{!}
	{
	\begin{tabular}{rr}
		\maxsizebox{\columnwidth}{!}
		{
			\tikzset{mark options={mark size=4, line width=1pt}}
			\begin{tikzpicture}
			\begin{axis}[%
			legend style={
				cells={anchor=west},
				legend pos={north west},
			},
			cycle list name={my color list},
			xmin=0, xmax=11,% x scale
			ymin=0, ymax=11, % y scale
			xlabel={Ranking  by $HPL$},
			ylabel={Ranking  by $HPCG$} ,
			scatter/classes={%
				MPP={ mark=diamond*,  draw=webgreen}}
			]
			\addplot[scatter,only marks,%
			scatter src=explicit symbolic]%
			table[meta=label] {
				x y label
				1	4	MPP
				2	2	MPP
				4   7   MPP
				5   6   MPP
				6   5   MPP
				7   3   MPP
				8   1   MPP
				9   10  MPP
				10   8   MPP
			};
			\addlegendentry{Rankings}

			\addplot+[ mark=diamond,  draw=webgreen] table[y={create col/linear regression={y=Y}},% mark=rectangle*,
			meta=label,    /pgf/number format/read comma as period
			] {
				x Y label
				1	4	MPP
				2	2	MPP
				4   7   MPP
				5   6   MPP
				6   5   MPP
				7   3   MPP
				8   1   MPP
				9   10  MPP
				10   8   MPP
			};
			\addlegendentry{Regression of rankings}
			
			\end{axis}
			\end{tikzpicture}
		}
		&
		\maxsizebox{\columnwidth}{!}
		{
			\tikzset{mark options={mark size=4, line width=1pt}}
			\begin{tikzpicture}
			\begin{axis}[%
			legend style={
				cells={anchor=west},
				legend pos={north west},
			},
			cycle list name={my color list},
			xmin=2e-8, xmax=3e-6,% x scale
			ymin=1e-5, ymax=3e-4, % y scale
			xlabel={$\alpha_{eff}$  by $HPL$},
			ylabel={$\alpha_{eff}$  by $HPCG$},
			xmode=log,
			ymode=log,
			scatter/classes={%
				MPP={ mark=diamond*,  draw=webgreen}}
			]
			\addplot[scatter,only marks,
			scatter src=explicit symbolic]%
			table[meta=label] {
				x y label
				3.273E-08	3.121e-5	MPP
				1.991E-07	2.882e-5	MPP
				9.656E-07   1.469e-4   MPP
				1.096E-07   3.910e-5   MPP
				1.590E-06   1.220e-4   MPP
				1.507E-06   6.092e-5   MPP
				1.040E-07   2.534e-5   MPP
				2.191E-07   7.353e-5  MPP
				1.221E-06   2.043e-4   MPP
			};
			\addlegendentry{$\alpha_{eff}$}

			\addplot+[ mark=diamond,  draw=webgreen] table[y={create col/linear regression={y=Y}},% mark=rectangle*,
			meta=label,    /pgf/number format/read comma as period
			] {
				x Y label
				3.273E-08	3.121e-5	MPP
				1.991E-07	2.882e-5	MPP
				9.656E-07   1.469e-4   MPP
				1.096E-07   3.910e-5   MPP
				1.590E-06   1.220e-4   MPP
				1.507E-06   6.092e-5   MPP
				1.040E-07   2.534e-5   MPP
				2.191E-07   7.353e-5  MPP
				1.221E-06   2.043e-4   MPP
			};
			\addlegendentry{Regression of $\alpha_{eff}$s}
			
			\end{axis}
			\end{tikzpicture}
		}
		\\
	\end{tabular}
}
	\caption{Correlation of ranking and $\alpha_{eff}$, derived using  $HPL$ and $HPCG$.}
	\label{fig:RankingVsAlpha}
\end{figure*}
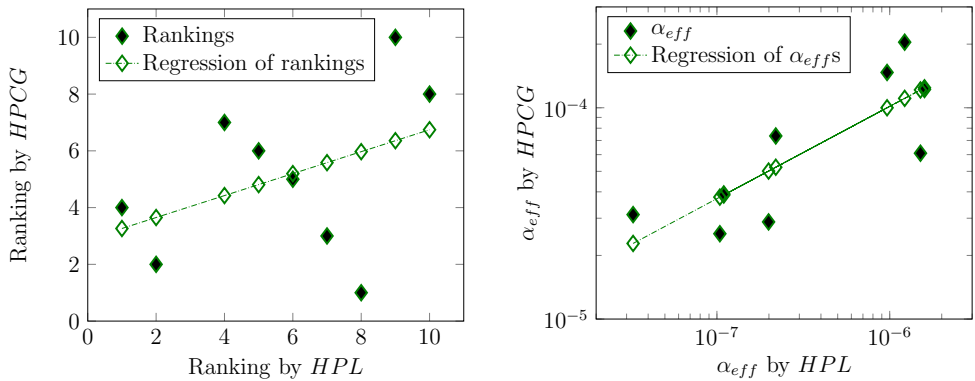

\subsection{Benchmarking supercomputers}

Benchmarks, utilized to derive numerical parameters for supercomputers, are specialized programs, which run on the \gls{HW}/\gls{OS} environment provided by supercomputer under test.
One can use benchmarks for different goals. Two typical fields of utilization:
to describe  the environment the supercomputer application runs in, and to guess
how quickly an application will run on a given supercomputer.

To understand operation of benchmarking, one needs to extend Amdahl's original model in such a way,
that non-parallelizable (i.e. apparently sequential) part comprises contributions from \gls{HW}, \gls{OS}, \gls{SW}\footnote{This separation cannot be strict: for example~\cite{FuSunwaySystem2016} utilizes 256+4 cores, where 4 cores assist system functionality at processor level, see~\cite{DongarraSunwaySystem:2016}.} and propagation time (\gls{PT}).
Among this, \gls{SW} represents what was assumed by Amdahl as the total sequential fraction.
As will be demonstrated in the discussion below, \textit{in the age of Amdahl, other contributions
	could be neglected compared to \gls{SW} contribution}.
Notice the different nature of the contributions.
They have only one common feature: \textit{they all consume time}.

Obviously, the (apparently) sequential fraction $(1-\alpha_{eff})$ cannot distinguish between
(at least apparently) sequential processing time contributions of different origin,
even the \gls{SW} (including \gls{OS}) and \gls{HW} contributions cannot be separated.
Similarly, it cannot be taken for sure that those contribution sum up linearly.
As long as computer components are in proximity in range $mm$,
contribution by \gls{PT} can be neglected.
Real-life supercomputer applications and the ones used for benchmarking differ very much in their contribution to the non-parallelizable fraction.
Different benchmarks provide different contributions to
the non-parallelizable fraction of the execution time, 
so comparing results derived using different benchmarks shall be done with maximum care.
\textit{Since the efficiency depends heavily on the number of cores, different configurations shall be compared using the same benchmark and same number of processors (or same $R_{Peak}$).}
Statements like "\textit{the benchmark results for the \gls{HPCG} benchmark reached only 0.3\% of 
	peak performance, which shows weaknesses of the architecture with slow memory and modest interconnect performance}", see~\cite{NSA-DOE_HPC_Report_2016},
clearly show that \gls{HPC} people are not aware of that in the case of having about two orders of magnitude higher number of processors \textit{Amdahl's law itself restricts performance} as experienced.
Utilizing benchmarks without considering this restriction would result in absolutely different rankings, both for \gls{HPL} and \gls{HPCG} benchmarks.

If the goal is to characterize the supercomputer's \gls{HW}+\gls{OS} system itself, a benchmark program should distort  \gls{HW}+\gls{OS} contribution as little as possible, i.e.
\gls{SW} contribution must be much lower than \gls{HW}+\gls{OS} contribution.
In the case of supercomputers, benchmark \gls{HPL} is used for this goal
since the beginning of the supercomputer age. 
The mathematical behavior of \gls{HPL} enables to 
minimize  \gls{SW} contribution.
The \gls{HW}+\gls{OS} contribution can also be lowered in such a way, as designers of Sunway do: 
they use 256+4 cores per processors, and 4 cores are dedicated to assist \gls{OS} functionality, see~\cite{DongarraSunwaySystem:2016}, thus reducing the duty of \gls{OS} to deal with cores by two orders of magnitude.
In this way the resulting non-parallelizable fraction  can be attributed 
mostly to \gls{HW}+\gls{OS} implementation.

If the goal is to estimate the expectable behavior of an application, 
the benchmark program should imitate the structure of the application;
i.e. to augment artificially the non-parallelizable fraction of the application to the amount typical in most supercomputer applications.
In the case of supercomputers, a couple of years ago \gls{HPCG} benchmark
has been introduced for this goal, since "\textit{\gls{HPCG} is designed to exercise computational and data access patterns that more closely match a different and broad set of important applications, and to give incentive to computer system designers to invest in capabilities that will have impact on the collective performance of these applications}", see~\cite{HPCG_List:2016}. 
However, its utilization can be misleading: \textit{the ranking is only valid for the \gls{HPCG} application, and only utilizing that number of processors.} This analysis reveals why supercomputer have two different rankings, see~\cite{DifferentBenchmarks:2017}, based on two different benchmarks: actually, $\alpha_{eff}^{SW}$ also contributes, so actually there are as many efficiencies (and rankings) as many benchmarks.

\subsection{Validating the $\alpha_{eff}$ model for supercomputers}\label{sec:validating}
To validate the $\alpha_{eff}$ model for supercomputers,  one can compare parameters and ranking 
derived using the \gls{HPL} and \gls{HPCG}, see~\cite{HPCG_List:2016} benchmarks, see Table~\ref{tab:benchmark}.
For the items in the table \gls{HW}/\gls{OS} environment (and so: the corresponding contributions to the non-parallelizable time) is the same,
the difference is caused by benchmark program structure.
The differences in efficiency values  delivered by the two benchmarks clearly show that efficiency differs by two orders of magnitude.
The $(1-\alpha_{eff})$ values give the explanation: \textit{the non-parallelizable fractions are 2-3 orders of magnitude higher when measured using \gls{HPCG} than when using \gls{HPL}}. 
It simply demonstrates a benchmarking artefact: an improper bechmark is used.

This helps to understand supercomputer development timeline: in \gls{HPL} approach, all contributions are 
decreased as much as possible. In \gls{HPCG} approach the \gls{SW} contribution dominates:
(all values $(1-\alpha_{eff}^{HPCG})$ are nearly equal, except those where $(1-\alpha_{eff}^{HPL})$ is an order of magnitude higher  because of the high \gls{HW} contribution),
and actually, the rest can be neglected.
According to Equ.~(\ref{eq:soverk}), the increased $(1-\alpha_{eff})$ value causes considerable differences in the efficiency.
For example, efficiencies of 'Cori' and 'Oakforest' differ by less than 10~\% when measured using \gls{HPL} and 
by more than 100~\% when measured using \gls{HPCG}.

\begin{table}
	\maxsizebox{\columnwidth}{!}
	{
		\begin{tabular} 
			{l|rrr|rrr} %
			\toprule
			\multicolumn{1}{c}{Computer Model}    & \multicolumn{3}{c}{HPL}& \multicolumn{3}{c}{HPCG}\\
			Top500 2017  & Efficiency & $(1-\alpha_{eff})$   & Rank  & Efficiency & $(1-\alpha_{eff})$   & Rank \\
			\midrule
			TaihuLight &	0.742 &	3.273E-08  & 1&	0.003 & 3.121e-5 & 4 \\
			Tianhe-2 & 0.617 & 1.991E-07 & 2 &	0.011& 2.882e-5 & 2 \\
			% 3 & Piz Daint & 361760 & 8.094E-07 & \bfseries{0.080}&	\bfseries{2.05E-08} \\
			Titan & 0.649  & 9.656E-07 & 4 & 0.012 & 1.469e-4&  7 \\ 
			Sequoia & 0.853 & 1.096E-07 & 5& 0.016 & 3.910e-5 & 6\\
			Cori & 0.503 & 1.590E-06 & 6 &0.013 & 1.220e-4 & 5 \\
			Oakforest& 0.544 & 1.507E-06 & 7& 0.028 & 6.092e-5 & 3 \\
			K computer & 0.932 & 1.040E-07 & 8 & 0.053 & 2.534e-5 & 1 \\
			Mira & 0.853 & 2.191E-07  & 9 &0.017 & 7.353e-5 & 10 \\
			Trinity & 0.731 & 1.221E-06  & 10 & 0.016 & 2.043e-4& 8 \\
			\bottomrule
		\end{tabular}
	}
	\caption{Comparing \gls{HPL} and \gls{HPCG}
		benchmark results of some TOP10 supercomputers}
	\label{tab:benchmark}
\end{table}

\begin{figure*}
	\begin{tikzpicture}[scale=.95]
	\begin{axis}
	[
	title={$R_{Max}$ of Taihulight supercomputer in function of $R_{Peak}$ with different benchmarks},
	width=\textwidth,
	cycle list name={my color list},
	legend style={
		cells={anchor=east},
		legend pos={north west},
	},
	xmin=1e-6, xmax=0.5,% x scale
	ymin=1e-6, ymax=0.5, % y scale
	xlabel={$R_{Peak}$  (exaFLOPS)},
	/pgf/number format/1000 sep={},
	ylabel={$R_{Max} (exaFLOPS)$},
	xmode=log,
	log basis x=10,
	ymode=log,
	log basis y=10,
	]
	\addplot table [x=a, y=h, col sep=comma] {RMaxvsRPeakatAlpha.csv};
	\addlegendentry{$1*10^{-8}$}
	\addplot table [x=a, y=h, col sep=comma] {RMaxvsRPeakatAlpha.csv};
	\addlegendentry{$HPL$}
	\addplot table [x=a, y=g, col sep=comma] {RMaxvsRPeakatAlpha.csv};
	\addlegendentry{$1*10^{-7}$}
	\addplot table [x=a, y=f, col sep=comma] {RMaxvsRPeakatAlpha.csv};
	\addlegendentry{$1*10^{-6}$}
	\addplot table [x=a, y=e, col sep=comma] {RMaxvsRPeakatAlpha.csv};
	\addlegendentry{$1*10^{-5}$}
	\addplot table [x=a, y=d, col sep=comma] {RMaxvsRPeakatAlpha.csv};
	\addlegendentry{$HPCG$}
	\addplot table [x=a, y=c, col sep=comma] {RMaxvsRPeakatAlpha.csv};
	\addlegendentry{$1*10^{-4}$}
	\addplot table [x=a, y=b, col sep=comma] {RMaxvsRPeakatAlpha.csv};
	\addlegendentry{$3*10^{-4}$}
	
	\addplot[only marks,  mark=o,  mark size=5, very thick] plot coordinates {
		(0.125,0.093) %Taihulight at HPL
		(0.125,0.000375) %Taihulight at HPCG
		(0.0113,0.0105) %K computer at HPL
		(0.0113,0.0006) %K computer at HPCG
	};
	\end{axis}
	\end{tikzpicture}
	\caption{$R_{Max}$ performance in function of peak performance $R_{Peak}$, at different $1-\alpha_{eff}()$ values.}
	\label{fig:exaRMaxAlpha}
\end{figure*}
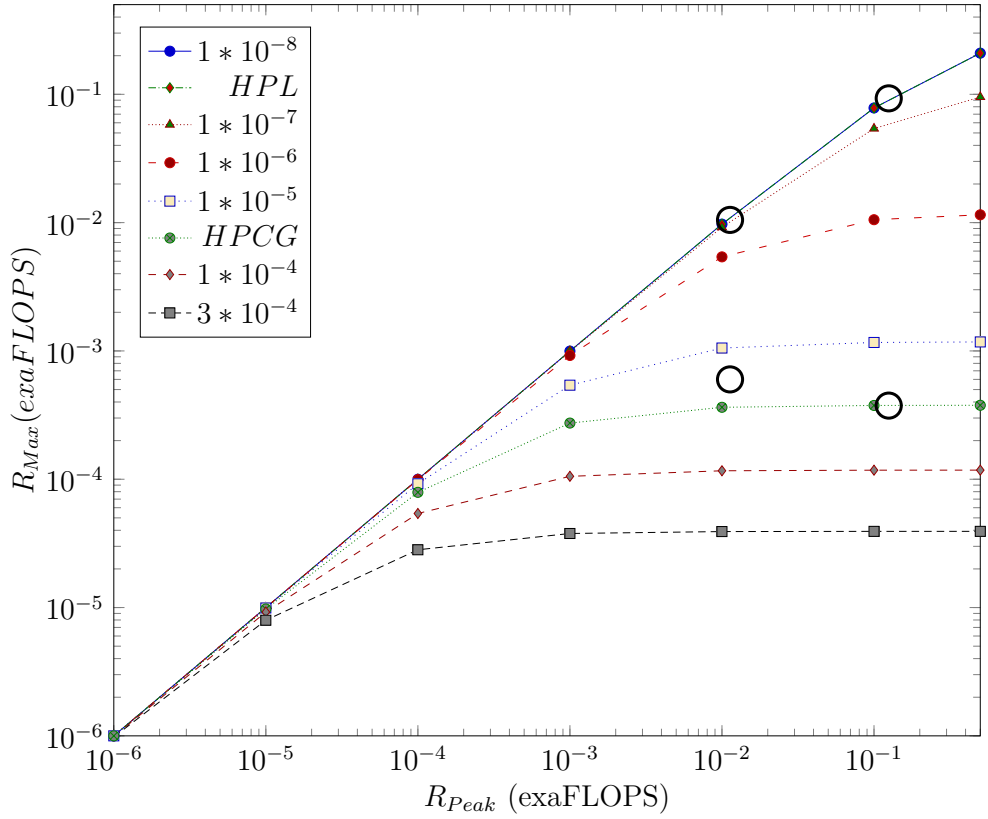

Benchmarks for a given supercomputer provide different $R_{Max}$, depending on value of
$(1-\alpha_{eff}^{SW})$ of the benchmark.
Fig.~\ref{fig:exaRMaxAlpha} displays some typical results, using Taihulight as an example.
(For other supercomputers the graphs are similar, except that the same $R_{Peak}$ is produced using processors with different single-processor performance, and so because of the different number of processors, the efficiency also changes.)
Bubbles on the figure mark positions of measured  $R_{Max}$ values, when using $HPL$ and $HPCG$ benchmarks, for  $Taihulight$ and $K\ computer$, respectively.
As noted above, in the case of \gls{HPCG} contribution of \gls{SW} "pulls down" the payload performance much stronger in the case of $Taihulight$ having much smaller $(1-\alpha_{eff}^{HPL})$, than in the case of $K\ computer$. 

As shown, building even bigger supercomputers has only sense for \gls{HPL}
class applications. 
\gls{HPCG} class applications enjoy just marginally better payload performance
than they would do on the same architecture equipped with an order of magnitude less processors.
Fig.~\ref{fig:exaRMaxAlpha} also helps to find out the optimum size of supercomputer for a specific application.

In the light of this discussion, \textit{results derived utilizing \gls{HPCG}
	as benchmark program give no direct information about hardware/software environment of the supercomputer}:
the contribution due to program structure (the measuring device) is at least two orders of magnitude higher than the contribution due to the  \gls{HW}/\gls{SW} system (the device under test). The roles are exchanged, and so the provided 'ranking' is the result of a kind of round-off effects, see the lack of correlation between the two rankings in Table~\ref{tab:benchmark}
and also in Fig.~\ref{fig:RankingVsAlpha}. (BTW: $(1-\alpha_{eff})$ strongly correlates with ranking, both for \gls{HPL} and \gls{HPCG}.)

Adopting \gls{HPCG} as benchmark enables hardware designers to utilize less expensive (and at the same time: less performable) architectural solutions,
and the resulting \gls{HW} will not be able to run \gls{HPL}  benchmark at some reasonable speed and/or efficiency, because in that case the relatively high \gls{HW}
contribution will dominate.
The same statement from another point of view:
"\textit{Designing a system for good \gls{HPL} performance can actually lead to
	design choices that are wrong for the real application mix, or add
	unnecessary components or complexity to the system}"~\cite{DongarraExascaleRace:2017}.
\textit{In the case of \gls{HPCG}, however, an order of magnitude higher \gls{OS} contribution or \gls{HW} contribution make no difference, and  this benchmark does not honour developments in this direction, so the development of supercomputers will follow a different path when directed by \gls{HPCG} as benchmark.
	Developing technology directed by \gls{HPCG} leads quickly and surely to dead-end street for high number of processors, see Fig.~\ref{fig:exaRMaxAlpha}, especially if 
	Exascale \gls{HPC} is targeted.}

The typical efficiency of TOP supercomputers is about 1~\% for benchmark \gls{HPCG}.
\textit{This means that supercomputers running application of class \gls{HPCG} make about 3-4~days payload work in a year.} As discussed above, contribution $\alpha_{eff}^{SW}$ is very different 
for different real-world applications, and because of this, \gls{HPCG} provides no real hints for the performance (\gls{HPL} provides even less: it describes \gls{HW}+\gls{OS}).
Measuring, however, $(1-\alpha_{eff})$ and looking at graphs in Fig.~\ref{fig:exaRMax} enables one to find the optimum $R_{Peak}$ for optimum performance/price.
%Paying 10 times more for 3~\% increase in performance is probable not a good business.
At this point knowing values of contributions $\alpha_{eff}^{HW+OS}$ and $\alpha_{eff}^{SW}$ would be a real advantage.
At least approximately, the first one describes the supercomputer, and the second one the application.

Just note that some dedicated measurements would enable to provide better estimations for the contributions:
making several dry (i.e. using the correct time but making no action) system handling calls
the slope of the (by the present model, linear) dependency of $(1-\alpha_{eff})$ on the number of system calls would provide the contribution of \gls{OS},
while the intercept delivers the (\gls{HW}+\gls{SW}) fraction.
Through changing the payload execution time, the \gls{SW} contribution can similarly be estimated, which finally enables to estimate also \gls{HW} contribution, which would be a real merit of \gls{HW}.

\subsection{Applications}  
In the case of 
benchmark programs  $(1-\alpha_{eff})$ is much lower, than 
in the case of real-life programs. (\cite{AmdahlSingleProcessor67} estimated the non-parallelizable part to be above 20~\%),
i.e. about 2-3 orders of magnitude higher than even that of
the \gls{HPCG} benchmark\footnote{Just recall that the goal of \gls{HPCG} is to provide a benchmark which behaves akin a typical \gls{HPC} application, rather than to characterize the supercomputer \gls{HW}+\gls{OS} system.}.
Because of this, \textit{efficiency of real-life programs decreases even more strongly with number of processors,
	so it is really worth to consider how many processors will provide optimum performance/price}. 

Supercomputer applications must be tuned to be scalable with number of processors.
For this goal $\alpha_{eff}$ can be utilized excellently.
Before and after some change in the program structure
the execution time can be measured using two different number of processors, and from those times  $\alpha_{eff}$ can be derived.
On the other side, using $\frac{R_{Max}}{R_{Peak}}$ of  a particular supercomputer (identical with $E$ in Equ.~(\ref{eq:soverk})) one can conclude efficiency (i.e. execution time) of a particular application using that supercomputer. Practically this would be the goal of utilizing \gls{HPCG}.

\section{Bounds on computing growth}

"\emph{The nature of this overhead (in
	parallelism) appears to be sequential so that it is unlikely to be amenable to
	parallel processing techniques. Overhead alone would then place an upper
	limit on throughput \dots, even
	if the housekeeping were done in a separate processor}~\cite{AmdahlSingleProcessor67}.

\subsection{A new exponential law of computing growth?}

In a recently published paper~\cite{ExponentialLawsComputing:2017} have pointed out that in real systems
the curves like Moore's law describing component density
(so, maybe also the development of $\alpha_{eff}$ with time, see Fig.~\ref{SupercomputerTimeline}),
will sooner or later saturate.
Their analyzis reveals that the exponential nature of the growth is the result of an interplay of many factors,
and also that such growth is better described by a "logistic curve"
which saturates after reaching some point.

Moore's law is formulated for several dependencies. 
In that case  a saturation point already reached: since cca. 2005 no more transistors can be added 
to a \gls{CPU} in a reasonable way.
For \textit{number of transistors in a chip} Moore's law still persists,
but not any more for \textit{number of transistors in a processor}.
Similar turning point was experienced with the manufacturing 
technology (lithograhic size) and is expected to occur with the atomic
nature of technological materials.
The saturation value, however, is not yet known and is different for the different reasons.

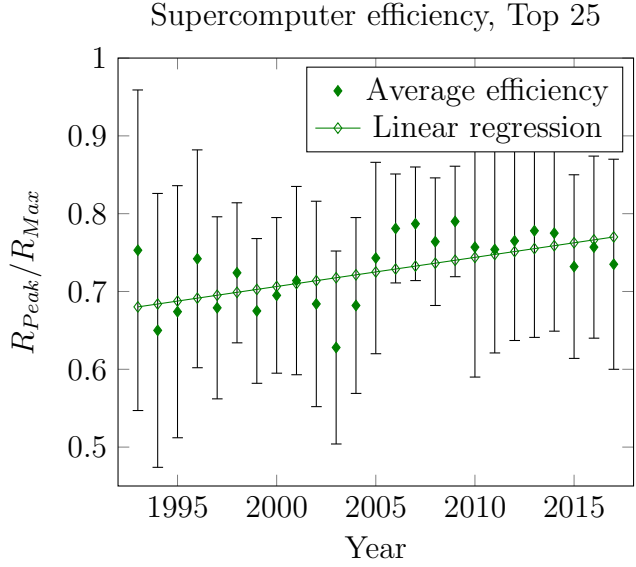
\begin{figure}
	\begin{tikzpicture}
	\begin{axis}
	[
	title={Supercomputer efficiency, Top 25},
	xmin=1992, xmax=2018,% x scale
	ymin=.45, ymax=1, % y scale
	xlabel=Year,
	/pgf/number format/1000 sep={},
	ylabel=$R_{Peak}/R_{Max}$,
	]
	\addplot [scatter,
	only marks,
	scatter src=explicit symbolic,
	scatter/classes={
		A0={ mark=diamond*, color=webgreen,
			/pgfplots/error bars/.cd,
			error mark options={draw=red}}},
	%		xmin=1993, xmax=2017,% x scale
	%		ymin=0.5, ymax=1.0, % y scale
	error bars/.cd,
	y dir=both,
	y explicit] 
	table[x=x,y=y,y error=err,meta=class,row sep=crcr] {
		x y err class\\
		2017 0.735 0.135   A0\\
		2016 0.757 0.117   A0\\
		2015 0.732 0.118   A0\\              
		2014 0.775 0.126   A0\\              
		2013 0.778 0.137   A0\\              
		2012 0.765 0.128   A0\\              
		2011 0.754 0.133   A0\\              
		2010 0.757 0.167   A0\\              
		2009 0.790 0.071   A0\\              
		2008 0.764 0.082   A0\\              
		2007 0.787 0.073   A0\\              
		2006 0.781 0.070   A0\\              
		2005 0.743 0.123   A0\\              
		2004 0.682 0.113   A0\\              
		2003 0.628 0.124   A0\\              
		2002 0.684 0.132   A0\\              
		2001 0.714 0.121   A0\\              
		2000 0.695 0.100   A0\\              
		1999 0.675 0.093   A0\\              
		1998 0.724 0.090   A0\\              
		1997 0.679 0.117   A0\\              
		1996 0.742 0.140   A0\\              
		1995 0.674 0.162   A0\\              
		1994 0.650 0.176   A0\\              
		1993 0.753 0.206   A0\\              
	};
	\addlegendentry{%
		Average efficiency} %
	\addplot+[mark=diamond, color=webgreen] table[row sep=\\, mark=triangle*,
	y={create col/linear regression={y=Y}}] % compute a linear regression from the
	%input table
	{
		x Y err class\\
		2017 0.735 0.135   A0\\
		2016 0.757 0.117   A0\\
		2015 0.732 0.118   A0\\              
		2014 0.775 0.126   A0\\              
		2013 0.778 0.137   A0\\              
		2012 0.765 0.128   A0\\              
		2011 0.754 0.133   A0\\              
		2010 0.757 0.167   A0\\              
		2009 0.790 0.071   A0\\              
		2008 0.764 0.082   A0\\              
		2007 0.787 0.073   A0\\              
		2006 0.781 0.070   A0\\              
		2005 0.743 0.123   A0\\              
		2004 0.682 0.113   A0\\              
		2003 0.628 0.124   A0\\              
		2002 0.684 0.132   A0\\              
		2001 0.714 0.121   A0\\              
		2000 0.695 0.100   A0\\              
		1999 0.675 0.093   A0\\              
		1998 0.724 0.090   A0\\              
		1997 0.679 0.117   A0\\              
		1996 0.742 0.140   A0\\              
		1995 0.674 0.162   A0\\              
		1994 0.650 0.176   A0\\              
		1993 0.753 0.206   A0\\              
	};
	\addlegendentry{%
		%        $\pgfmathprintnumber{\pgfplotstableregressiona} \cdot x
		%        \pgfmathprintnumber[print sign]{\pgfplotstableregressionb}$ 
		Linear regression} %
	\end{axis}
	\end{tikzpicture}
	\caption{Trend of development of average value of ($\frac{R_{Max}}{R_{Peak}}$) in the past 25 years, calculated for the first 25 supercomputers in the year in question.  Data derived using the \gls{HPL} benchmark.}
	\label{SupercomputerEfficiencyAverage}
\end{figure}

%\input{fig/EfficiencyDependenceCompare}
%\MEtikzfigure[wide]{
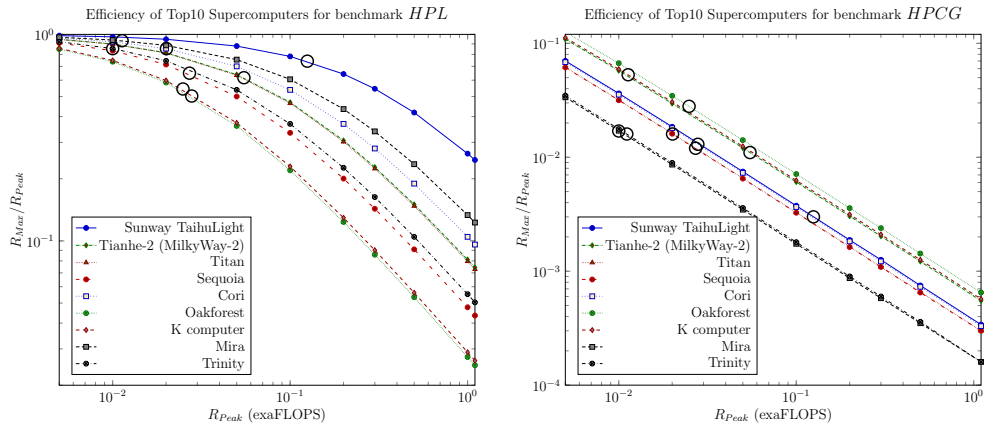
\begin{figure*}
	\maxsizebox{\textwidth}{!}{
	\begin{tabular}{cc}
		\begin{tikzpicture}[scale=.5]
		\begin{axis}
		[
		title={Efficiency of Top10 Supercomputers for benchmark \large $HPL$},
		width=\textwidth,
		cycle list name={my color list},
		legend style={
			cells={anchor=east},
			legend pos={south west},
		},
		xmin=0.005, xmax=1.1,% x scale
		ymin=2e-2, ymax=1, % y scale
		xlabel={$R_{Peak}$  (exaFLOPS)},
		/pgf/number format/1000 sep={},
		ylabel={$R_{Max}$/{$R_{Peak}$}},
		xmode=log,
		log basis x=10,
		ymode=log,
		log basis y=10,
		]
		\addplot table [x=a, y=b, col sep=comma] {EffHPL10.csv};
		\addlegendentry{Sunway TaihuLight}
		\addplot table [x=a, y=c, col sep=comma] {EffHPL10.csv};
		\addlegendentry{Tianhe-2 (MilkyWay-2)}
		\addplot table [x=a, y=d, col sep=comma] {EffHPL10.csv};
		\addlegendentry{Titan}
		\addplot table [x=a, y=e, col sep=comma] {EffHPL10.csv};
		\addlegendentry{Sequoia}
		\addplot table [x=a, y=f, col sep=comma] {EffHPL10.csv};
		\addlegendentry{Cori}
		\addplot table [x=a, y=g, col sep=comma] {EffHPL10.csv};
		\addlegendentry{Oakforest}
		\addplot table [x=a, y=h, col sep=comma] {EffHPL10.csv};
		\addlegendentry{K computer}
		\addplot table [x=a, y=i, col sep=comma] {EffHPL10.csv};
		\addlegendentry{Mira}
		\addplot table [x=a, y=j, col sep=comma] {EffHPL10.csv};
		\addlegendentry{Trinity}
		\addplot[only marks,  mark=o,  mark size=5, very thick] plot coordinates {
			(0.125,0.742) %Taihulight
			(0.0549,0.617) %Tianhe-2
			(0.0271,0.649) %Titan
			(0.0201,0.853) %Sequoia
			(0.0279,0.503) %Cori
			(0.0249,0.544) %Oakforest
			(0.0113,0.932) %K computer
			(0.0100,0.853) %Mira
			(0.0111,0.016) %Trinity
		};
		\end{axis}
		\end{tikzpicture}
		&
		\begin{tikzpicture}[scale=.5]
		\begin{axis}
		[
		title={Efficiency of Top10 Supercomputers for benchmark \large $HPCG$},
		width=\textwidth,
		cycle list name={my color list},
		legend style={
			cells={anchor=east},
			legend pos={south west},
		},
		xmin=0.005, xmax=1.1,% x scale
		ymin=1e-4, ymax=0.12, % y scale
		xlabel={$R_{Peak}$ (exaFLOPS)},
		/pgf/number format/1000 sep={},
		ylabel={$R_{Max}$/{$R_{Peak}$}},
		xmode=log,
		log basis x=10,
		ymode=log,
		log basis y=10,
		]
		\addplot table [x=a, y=b, col sep=comma] {EffHPCG10.csv};
		\addlegendentry{Sunway TaihuLight}
		\addplot table [x=a, y=c, col sep=comma] {EffHPCG10.csv};
		\addlegendentry{Tianhe-2 (MilkyWay-2)}
		\addplot table [x=a, y=d, col sep=comma] {EffHPCG10.csv};
		\addlegendentry{Titan}
		\addplot table [x=a, y=e, col sep=comma] {EffHPCG10.csv};
		\addlegendentry{Sequoia}
		\addplot table [x=a, y=f, col sep=comma] {EffHPCG10.csv};
		\addlegendentry{Cori}
		\addplot table [x=a, y=g, col sep=comma] {EffHPCG10.csv};
		\addlegendentry{Oakforest}
		\addplot table [x=a, y=h, col sep=comma] {EffHPCG10.csv};
		\addlegendentry{K computer}
		\addplot table [x=a, y=i, col sep=comma] {EffHPCG10.csv};
		\addlegendentry{Mira}
		\addplot table [x=a, y=j, col sep=comma] {EffHPCG10.csv};
		\addlegendentry{Trinity}
		\addplot[only marks,  mark=o,  mark size=5, very thick] plot coordinates {
			(0.125,0.003) %Taihulight
			(0.0549,0.011) %Tianhe-2
			(0.0271,0.012) %Titan
			(0.0201,0.016) %Sequoia
			(0.0279,0.013) %Cori
			(0.0249,0.028) %Oakforest
			(0.0113,0.053) %K computer
			(0.0100,0.017) %Mira
			(0.0111,0.016) %Trinity
		};
		\end{axis}
		\end{tikzpicture}
		\\
	\end{tabular}
}
	\caption{Comparing efficiencies of selected TOP10 (as of 2017 July) supercomputers
		in function of their peak performance $R_{Peak}$, for \gls{HPL} and \gls{HPCG} benchmarks. The actual $R_{Peak}$ values are denoted by a bubble.}
	\label{fig:exaefficiency}
\end{figure*}

\subsection{Limitations of building larger supercomputers}

Fig.~\ref{SupercomputerTimeline} shows up a behavior very similar to that of Moore's law,
(it looks like that  $(1-\alpha_{eff})$ decreases year-by-year by a factor of cca. $1.5$).
Some reasons can also be seen why also this behavior 
is not without limitations.
Using some reasonable assumptions about the different contributions
to the non-parallelizable fraction mentioned above,
their order of magnitude can be estimated, 
and also some saturation values can be forecasted.
It is interesting to note that in the \textit{"Summary Report of the 
	Advanced Scientific Computing Advisory Committee (ASCAC) Subcommittee"}, see~\cite{ScienceExascaleRace:2010}, \textit{the feasibility
	of building supercomputers of arbitrary size}  (in the sense 
whether any physical law or computer science principle restricts the achievable computing throughput)
\textit{remained out of scope.} Quite similarly, the European Union's Action Plan, see~\cite{EUActionPlan:2016}, 
assumes no limitations, rather that
"\textit{With a differentiated strategy and sufficient investment and political will, Europe can be
	a global player in \gls{HPC}}".

For calculating bounds based on our extended model data published
by~\cite{DongarraSunwaySystem:2016} are used.
The 13,298 seconds benchmark runtime on the 1.45 GHz processors means 
$2*10^{13}$ clock periods. The absolutely necessary non-parallelizable activity is to start and stop the calculation. 
If starting and stopping a zero-sized supercomputer without \gls{OS} could be done in 2 clock periods,
then the absolute limit for $(1-\alpha_{eff})$ would be
$10^{-13}$.

If one considers a cca. 100~meter sized computer having 1~GHz cores, the signal round trip time is cca. 
$10^{-6}$ seconds, or $10^{3}$ clock periods, and a network message can be estimated to be of length $10^{-5}$ seconds (including operating time of HW), or $10^{4}$
clock periods.
So, the absolute limit for $(1-\alpha_{eff})$ of a computer with
realistic size, but no operating system is $10^{-9}$.

We need to use, however, an operating system. If one considers context change with its consumed $10^{4}$ cycles, the absolute limit is cca. $10^{-9}$,
on a zero-sized supercomputer.
\textit{In addition, millions of cores must be manipulated
	through the system call, which contribution increases linearly with the number of cores
	and contribution from \gls{OS} can be dominant at high number of cores.} This is why designers of Sunway dedicated 4 cores per processor~\cite{FuSunwaySystem2016} to reduce this dependence by two orders of magnitude.
As discussed, 
the application itself produces some non-payload activity,
which can be assumed also to be at least in the range of $10^{4}$-$10^{5}$  clock cycles. 

It is probably a realistic estimation, that contributions of \gls{HW}, \gls{OS}, \gls{PT} and \gls{SW} (the application itself)
can sum up to  at least $10^{5}$ clock cycles, resulting in a $10^{-8}$
absolute limit for $(1-\alpha_{eff})$. 
Although it is a very rough
estimation, it is worth to compare it to the value  $3.3*10^{-8}$, calculated from  data published by~\cite{DongarraSunwaySystem:2016} for the Taihulight supercomputer, see~\cite{FuSunwaySystem2016},
(and also shown in Fig.~\ref{SupercomputerTimeline}). %Absolutely the same order of magnitude. 

As it is known from textbooks, according to Amdahl's law, the available maximum speedup
(the apparent computing throughput) is given by $\frac{1}{1-\alpha}$,
i.e. for the derived limiting value is about $10^8$.
This should be multiplied with the computing throughput of a typical processor used 
in supercomputers, typically 10~$Gflop/s$.
This results in $10^{18}~flop/s$, i.e. about the "dream limit",
targeted by several supercomputer building teams. 
To increase this product, either the single-processor
performance must be increased or $(1-\alpha)$ decreased, or both.
In the case of benchmark \gls{HPCG} $\frac{1}{1-\alpha}$ is in the order of magnitude $10^{-5}$ (see Table~\ref{tab:benchmark}), and correspondingly the upper limit for computing throughput is 0.001~Eflop/s. This means that for the "\textit{broad set of important applications}"~\cite{HPCG_List:2016} the "dream limit" cannot be achieved at all,
and even for the much less broad set of applications of class  \gls{HPL} it is questionable.

\subsection{A new experimental evidence}
A nice evidence was provided by the new supercomputer appearing
in the 2017 November list,
that the theoretical bound achieved. As it could be quessed,
the constructors bought 20M processors to build the new \#1 supercomputer,
with \textit{nominal} performance $R_{Max}=229~Pflop/s$, twice more than that of Taihulight.
The provided benchmark data ($E=0.679$ and $R_{Max}= 19.136~Pflop/s$, measured using 2.4M cores only), however, qualified them to catch the 4th position only.
If they could keep the same efficiency and would use all cores, 
they should have $R_{Max}= 155.3~Pflop/s$, a new world record.
It is a big question, if "\textit{The system's 19,860,000 cores represent the highest level of concurrency ever recorded on the TOP500 rankings of supercomputers}", see~\cite{Gyoukou:2017}, why did they participate in the race
with using only 12~\% of the cores. Why they did not want to be the \#1?

The analysis above gives the answer:
even if we assume that effective parallelism will not be worse due to the higher number of processors, the efficiency would drop to $E_{10M}=0.21$,
and this efficiency would drop out from the average value cca. $0.73$ 
calculated for the TOP25 supercomputers, see Fig.~\ref{SupercomputerEfficiencyAverage}.  Even with this low efficency, 
$R_{Max}=48~Pflop/s$ could be produced, qualifying them to catch the 2nd position.
However, increasing the number of cores increases the looping delay
and causes to increase magnitude of value of $(1-\alpha_{eff,10M})$.
The measurement conditions are not known, so  the amount of increase can only be guessed. 
If  $(1-\alpha_{eff})$ would increase by a factor of 2, Gyoukou  could take position 3 with 
$E_{10M}=0.116$ and $R_{Max}= 26.56~Pflop/s$. 
If it would increase by a factor of 5, Gyoukou  could take position 10 with 
$E_{10M}=0.05$ and $R_{Max}= 11.45~Pflop/s$. 
If it would increase by a factor of 8, Gyoukou  could take position 14 with 
$E_{10M}=0.032$ and $R_{Max}= 7.38~Pflop/s$. 

As discussed, $(1-\alpha_{eff}^{HPCG})$ is about two orders higher than $(1-\alpha_{eff}^{HPL})$. Since  values $(1-\alpha_{eff}^{HPL})$ are nearly the same for Tianhe-2 (MilkyWay-2) in the 2nd position and Gyoukou in the 4th position, and also the number of processors are approximately the same, one can assume 
$(1-\alpha_{eff, 2.4M}^{HPCG})= 3*10^{-5}$ for Gyoukou.
Using assumptions similar to the ones used for \gls{HPL}, $E^{10M}=0.0017$, $R_{Max}^{HPCG,10M}= 0.39~Pflop/s$.
Oakforest at the 9th position also has $R_{Max}^{HPCG,0.56M}= 0.385~Pflop/s$,
i.e. nearly the same payload performance, but using only 35 times less processors
or 9 times less nominal performance. That is: less is more, because of Amdahl's law. 

This analyzis provokes the question: \textit{how much is it realistic to plan building even larger \gls{SPA} supercomputers}?

 \subsection{Forecasts for exaFLOPS supercomputers}

 \begin{table}
 	\maxsizebox{\columnwidth}{!}
 	{
 		\begin{tabular} 
 			{rlrrrr} %
 			TOP500  & Computer Model    & N proc  & ($1-\alpha_{eff}$) & Efficiency at  & ($1-\alpha_{eff}$) at\\
 			rank 2017  &  & in 2017   &  & 1~exaFLOPS & 1~exaFLOPS\\
 			\midrule
 			1 & %Sunway 
 			TaihuLight &	10649600 &	3.273E-08  & \bfseries{0.265}&	\bfseries{4.11E-09} \\
 			2 & Tianhe-2& 3120000 & 1.991E-07 & \bfseries{0.081}&	\bfseries{1.09E-08} \\
 			3 & Piz Daint & 361760 & 8.094E-07 & \bfseries{0.080}&	\bfseries{2.05E-08} \\
 			4 & Titan & 560640 & 9.656E-07 & \bfseries{0.048}&	\bfseries{2.62E-08} \\ 
 			5 & Sequoia & 1572864 & 1.096E-07 & \bfseries{0.105}&	\bfseries{2.21E-09} \\
 			6 & Cori & 622336 & 1.590E-06 & \bfseries{0.027}&	\bfseries{4.43E-08} \\
 			7 & Oakforest-PACS & 556104 & 1.507E-06 & \bfseries{0.029}&	\bfseries{3.75E-08} \\
 			8 & K computer & 705024 & 1.040E-07 & \bfseries{0.133}&	\bfseries{1.17E-09} \\
 			9 & Mira & 786432 & 2.191E-07  & \bfseries{0.055}&\bfseries{2.21E-09} \\
 			10 & Trinity & 301056 & 1.221E-06  & \bfseries{0.029}&\bfseries{1.35E-08}\\
 		\end{tabular}
 	}
 	\caption{$\frac{R_{Peak}}{R_{Max}}$ of present TOP10 supercomputer architectures upgraded with more cores to provide 1~exaFLOPS, or the $(1-\alpha_{eff})$ to be achieved to keep their present efficiency.  Data are derived using the \gls{HPL} benchmark. }
 	\label{tab:exaflop}
 \end{table}

 %\MEtikzfigure[wide]{
 \begin{figure*}
 	{
 		\begin{tikzpicture}[scale=.95]
 		\begin{axis}
 		[
 		title={$R_{Max}$ of Top10 Supercomputers for benchmark \huge $HPL$},
 		width=\textwidth,
 		cycle list name={my color list},
 		legend style={
 			cells={anchor=east},
 			legend pos={north west},
 		},
 		xmin=0.005, xmax=1.1,% x scale
 		ymin=0.003, ymax=0.3, % y scale
 		xlabel={$R_{Peak}$  (exaFLOPS)},
 		/pgf/number format/1000 sep={},
 		ylabel={$R_{Max} (exaFLOPS)$},
 		xmode=log,
 		log basis x=10,
 		ymode=log,
 		log basis y=10,
 		]
 		\addplot table [x=a, y=b, col sep=comma] {RMaxHPL10.csv};
 		\addlegendentry{Sunway TaihuLight}
 		\addplot table [x=a, y=c, col sep=comma] {RMaxHPL10.csv};
 		\addlegendentry{Tianhe-2 (MilkyWay-2)}
 		\addplot table [x=a, y=d, col sep=comma] {RMaxHPL10.csv};
 		\addlegendentry{Titan}
 		\addplot table [x=a, y=e, col sep=comma] {RMaxHPL10.csv};
 		\addlegendentry{Sequoia}
 		\addplot table [x=a, y=f, col sep=comma] {RMaxHPL10.csv};
 		\addlegendentry{Cori}
 		\addplot table [x=a, y=g, col sep=comma] {RMaxHPL10.csv};
 		\addlegendentry{Oakforest}
 		\addplot table [x=a, y=h, col sep=comma] {RMaxHPL10.csv};
 		\addlegendentry{K computer}
 		\addplot table [x=a, y=i, col sep=comma] {RMaxHPL10.csv};
 		\addlegendentry{Mira}
 		\addplot table [x=a, y=j, col sep=comma] {RMaxHPL10.csv};
 		\addlegendentry{Trinity}
 		\addplot[only marks,  mark=o,  mark size=5, very thick] plot coordinates {
 			(0.125,0.09275) %Taihulight
 			(0.0549,0.033873) %Tianhe-2
 			(0.0271,0.01759) %Titan
 			(0.0201,0.01715) %Sequoia
 			(0.0279,0.01403) %Cori
 			(0.0249,0.01355) %Oakforest
 			(0.0113,0.01053) %K computer
 			(0.0100,0.00853) %Mira
 			(0.0111,0.00178) %Trinity
 		};
 		\end{axis}
 		\end{tikzpicture}
 		
 		\caption{$R_{Max}$ performance of selected TOP10 (as of 2017 July) supercomputers
 			in function of their peak performance $R_{Peak}$, for the \gls{HPL} benchmark. The actual $R_{Peak}$ values are denoted by a bubble.}
 		\label{fig:exaRMax}
 	}
 \end{figure*}
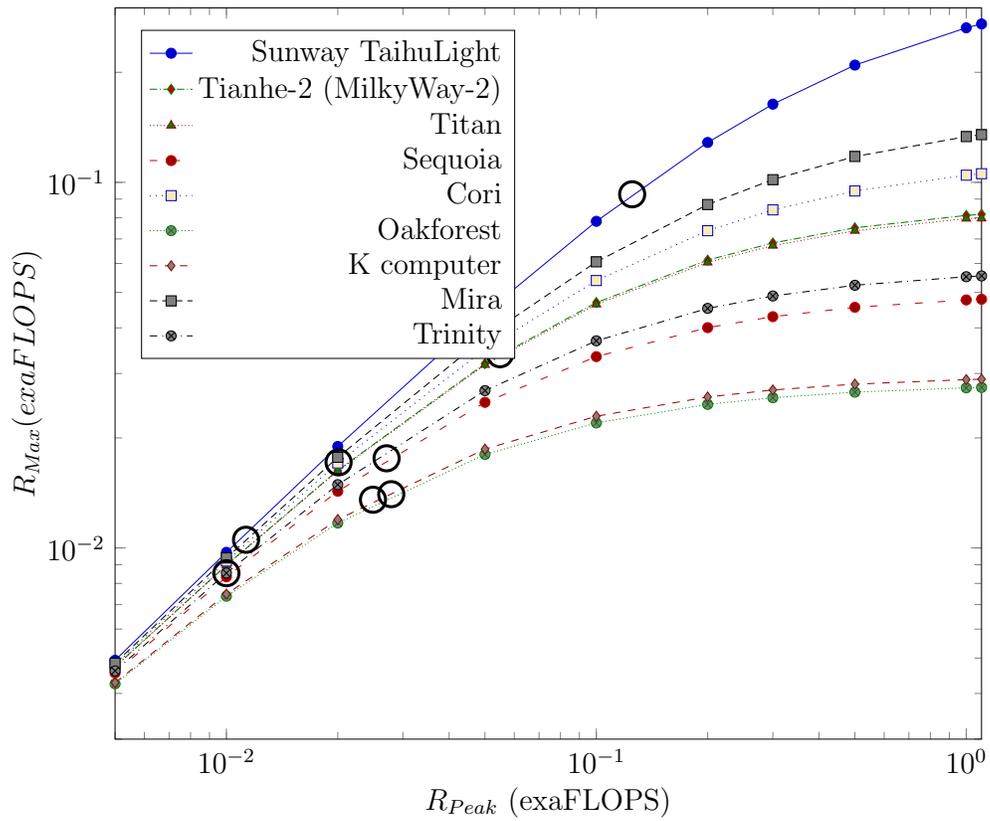

 From TOP500 data $R_Max$ values in function of $R_{Peak}$ can be calculated, see
 Fig.~\ref{fig:exaefficiency} (i.e. virtually the number of processors is changed for the different configurations).
 Since the efficiency values differ by orders of magnitude, the behavior of efficiencies
 changes drastically between the two benchmarks.
 The reported (measured) efficiency values are marked by bubbles on the figures.

 The pre-last column of Table~\ref{tab:exaflop} displays the  
 $\frac{R_{Max}}{R_{Peak}}$ value for the first ten (as of 2017 July) supercomputers 
 on the TOP500 list, if they were upgraded with more cores to provide 1~exaFLOPS.
 (For the trend of $HPL$ efficiency see Fig.~\ref{SupercomputerEfficiencyAverage}.
 Probably, there is not much sense to build supercomputers with efficiency 
 below one percent, but the fact that benchmark \gls{HPCG} is accepted 
 as a new metric for ranking supercomputer systems, seems to contradict to it.)
 
 Fig.~\ref{fig:exaefficiency} also underlines importance of benchmarking:
 benchmark \gls{HPCG} produces not only drastically lower efficiency values, the nature of
 efficiency also changes: while on the left side increasing ${R_{Peak}}$ by two orders of magnitude
 triggers \textit{less} than two orders of magnitude  decrease in efficiency,
 on the right side the efficiency decreases \textit{more} than two orders of magnitude.
 Similarly to the case shown in Table~\ref{tab:Dongarra1992}: putting more processors in an architecture,
 without making efforts for enhancing  $\alpha_{eff}$ results in tragically low efficiency.
 
 %Building exaFLOPS supercomputers for applications akin \gls{HPCG} is nonsense: 
 %the more processors, the less efficiency, thanks to the two orders of magnitude higher $(1-\alpha_{eff})$ value.
 %Supercomputers running \gls{HPCG} will provide just a few percent efficiency,
 %and provide $R_{Max}$ about 0.1~exaFLOPS. To achieve 1~exaFLOPS $R_{Max}$,
 %a few times $10^{-9}$ value of  $(1-\alpha_{eff})$ should be achieved
 
 Fig.~\ref{fig:exaRMax} shows absolute computing performance $R_{Max}$
 for some of TOP10 supercomputers. When calculating the diagram lines, 
 virtually the number of processors were varied. The actual value is 
 denoted by a bubble. As expected,  $R_{Max}$ for Taihulight seems
 to saturate around .35 exaFLOPS, for the rest of supercomputers at much lower values.

One can also calculate from reverting Equ.~(\ref{eq:soverk}) what enhancement in $(1-\alpha_{eff})$ would be 
necessary for those configurations to achieve 1 exaFLOPS and at the same time to keep their present $\frac{R_{Peak}}{R_{Max}}$, keep their present efficiency, see
the last column of Table~\ref{tab:exaflop}.
For deriving the results, benchmark \gls{HPL} was assumed.
As from Fig.~\ref{SupercomputerEfficiencyAverage} can be concluded, for future supercomputers achieving $\frac{R_{Max}}{R_{Peak}}$ around $0.73$
can be expected\footnote{Provided that \gls{HPL} benchmark will not be replaced by \gls{HPCG}}, the development will have to target reducing $(1-\alpha_{eff})$.

In the light of the analyzis above the primary candidates to deliver 1 exaFLOPS are the supercomputers 
which are able to produce presently  $(1-\alpha_{eff})$ around $10^{-8}$.
To achieve considerably lower 
$(1-\alpha_{eff})$, all contributions mentioned in our model must be considerably lowered.
The physical size (mainly due to need of cooling) can hardly be decreased,
although 3D arrangements can help also here.
The present layering of computing stack requires context change, starting(preparing)/stopping an application is inevitable and the linearly increasing contribution of  \gls{OS} due to the large number of cores must be eliminated.
Lowering only one of these contributions is useless.
\textit{That is, at least the physical size (speed of the light) really puts an upper bound to supercomputers performance.}

At least if utilizing Single Processor Approach persist. 	 Shall we also prepare for a post-Amdahl era?
As~\cite{MarkovLimitsOfLimits:2014} pointed out, in computing
the limitations are also limited. This holds also for supercomputing:
with introducing cooperating processors, see~\cite{VeghEMPAthY86:2016} and reasonable layering, see~\cite{VeghLayering:2017}, the landscape changes drastically.

\begin{flushright}

\end{flushright}
\subsection{Is perfectness the common reason?}

The number of similarities between the present law depicted in Fig.~\ref{SupercomputerTimeline} and Moore's law suggests to find some common reason.
An interesting idea is that in both cases a kind of "\textit{perfectness}" is approached.
The goal to be achieved in the case of Moore's law is infinitesimally small
component size, in the case of parallelization infinitesimally small non-parallelizable part.
Both dependencies show exponential behavior, and as presented, both of them will behave (sooner or later) as a logistic curve. 
These laws are able to forecast the expected behavior of performance in the coming years,
and serious consideration must be given to their scope of validity.

\section{Conclusion}

The paper validated that Amdahl's 50-years old model (with slight extension) correctly describes operation of parallelized computing systems, provided that the meaning of terms used in the model are properly interpreted.
Originally, Amdahl's law was interpreted for \gls{SW} contribution only, corresponding to that-time stage of technology: compared to the contribution of \gls{SW} to the 
non-parallelizable fraction of the model, the rest of contributions were negligible.
The development of technology (including different kinds of accelerators, appearance of networked communication inside and among chips, need for cooperation of several processors in forms of grids, \gls{MCP}s and  supercomputers, etc.) lead to forgetting this
universal law. 

The law has been reformulated, now explicitly giving time dimension to the terms.
It was also emphasized, that -- as suggested originally by Amdahl -- all contributions
which are not parallelizable, will appear as (at least apparently) sequential fraction.
Although those latter contributions are not summed up linearly (some sequential contributions
may be partly parallel with each other), these two main classes of contributions
can serve as a proper base for understanding parallelism in modern computer systems.

The case of analyzing load balancing compiler, see~\cite{Vegh:2017:AlphaEff}, is
closest to the original assumptions of Amdahl.
Contribution of \gls{SW} is in the order of several percents, \gls{HW}
does not contribute too much at those low number of cores, and \gls{OS} 
makes contribution which is low and not separable from that of \gls{SW}.

In the case of analyzing communication inside chip~\cite{Vegh:2017:AlphaEff} a mixture of \gls{HW}, \gls{SW} and \gls{OS} and 
propagation delay time contributes to the non-parallelizable fraction.
The latter one is not really significant inside chip, but might even be dominant
if analyzing systems connected through some kind of network,
with all subtleties of networks.
Here even non-parallelizable contributions are distributed among 
communicating cores, and because of this, resulting non-parallelizable fraction
is one order of magnitude lower than in the previous case. 

When analizing supercomputers having largely different number of processors,
the real nature of of Amdahl's law can be studied and understood.
In that case both the propagation delay (dozens of meters instead of several microns) 
and the number of processors (millions rather than dozens)
is bigger, by several orders on magnitude, than in the case of "normal" systems.
Using the excellent public database~\cite{Top500:2016}, the assumptions of the model could be thoroughly tested. At the beginning of the supercomputer age,
\gls{SW} dominated the non-parallelizable fraction.
The need for producing ever more effective assemblies of ever more processors
directed the development in such a way that \textit{all} contributions
had to be minimized to achieve reasonable efficiency.
This method of development covered that the initially dominant \gls{SW}
gradually gave place for the other contributions.
Although utilizing very sophisticated methods of \gls{HW} and \gls{SW} engineering
enabled to reduce nearly all contributions (and in principle further improvements are possible),
the propagation delay is limited by speed of the light: finite physical size of 
components and their energy dissipation. Because of this, until technology changes,
the maximum performance of supercomputers cannot break this theoretical barrier.

One can really conclude that \textit{Amdahl's law is really a very basic law of computer science
	and if its underlying model also considers the propagation delay,
	it correctly describes all experiences and issues of parallel operations,
	including the cutting edge supercomputing systems}.
The model provides surprisingly good numerical values for all published performance data,
although dedicated measurements are needed to pinpoint role, size and interaction of 
components in the suggested model.

Results about forecasting parameters of future supercomputers are especially important.
Amdahl's law, using model extended as described above, correctly forecasts
performance limitations, both of supercomputer applications and of supercomputer architecture itself. 
Using that model, the absolute performance bound of 
supercomputers was concluded, furthermore it was pointed out that serious enhancements are still necessary to achieve the exaFLOPS dream value.
\textit{One should notice again, as was also emphasized by Amdahl: conclusions
	and calculations are only valid for computers
	built in \gls{SPA}.}
Introducing new computing principles (or at least using less restrictive interpretation of the classic ones) may invalidate the contents of this chapter
and open a new chapter of computing.

%Amdahl's law (published in a quarter century before 
%supercomputing was born) seems to bound both supercomputer 
%applications and supercomputer architecture itself. 
%The introduced $\alpha_{eff}$ parameter, actually: the (at least apparently) sequential fraction  comprising contributions of different origin,
%can be used
%to describe architecture and operation of supercomputers
%built in Single Processor Approach, and can effectively assist both supercomputer application makers and supercomputer constructors.
%
%The simple model describes all experiences acquired by the supercomputer community. 
%The model provides surprisingly good numerical values for all published performance data,
%although dedicated measurements are needed to pinpoint role, size and interaction of the 
%components in the suggested model.
%The paper validates that Amdahl's 50-years old model (with slight extension) correctly describes the performance limitations
%of the present supercomputers. 
%Using some simple and reasonable assumptions,  the absolute performance bound of 
%supercomputers was concluded, furthermore that serious enhancements are still necessary to achieve the exaFLOPS dream value.

\section*{Acknowledgements}
Project no. 125547 has been implemented with the support provided from the National Research, Development and Innovation Fund of Hungary, financed under the K funding scheme.

\section*{References}

\bibliographystyle{elsarticle-harv} 
\bibliography{Bibliography}
\end{document}